\documentclass[fleqn,usenatbib]{rasti}

\usepackage{amsmath}
\usepackage{amssymb}
\usepackage{txfonts}
\usepackage{upgreek}
\usepackage[T1]{fontenc}
\usepackage{graphicx}
\usepackage{booktabs}
\usepackage{multirow}
\usepackage{xcolor}
\usepackage{tikz}
\usepackage{array}
\newcolumntype{P}[1]{>{\raggedright\arraybackslash}p{#1}}
\usetikzlibrary{arrows.meta,positioning,shapes.geometric,fit,backgrounds}
\hypersetup{colorlinks=true,citecolor=blue,linkcolor=blue,urlcolor=blue}

\graphicspath{{figures/}}

\newcommand{\um}{\ensuremath{\upmu}m}
\newcommand{\HRI}{HRI}
\newcommand{\ep}{e$^-$}

\newlength{\originalParindent}                                            
\newenvironment{my_itemize}
{
    \setlength{\originalParindent}{\parindent}                            
    \leftskip 3mm                                                        
    \parindent -0.11in

    \newcommand{\originalItem}{\item}                                     
    \renewcommand{\item}{$\bullet$~}                                      
}
{                                                                         
    \par

    \leftskip 0in
    \setlength{\parindent}{\originalParindent}                            
    \renewcommand{\item}{\originalParindent}                              
}

\title[A design concept for the HWO High Resolution Imager]
{A design concept for the High Resolution Imager\\ on the Habitable Worlds Observatory
}\author[Van Eylen et al.]{Vincent Van Eylen$^{1}$,
Richard Massey$^{2}$\thanks{E-mail: r.j.massey@durham.ac.uk},
Alan Smith$^{1}$,
Jerome Amiaux$^{3}$,
Florence Ardellier$^{3}$,
Saeeda Awan$^{1}$,
\newauthor
Patrick Curry$^{1}$,
Daniel Dicken$^{4}$,
Paul Eccleston$^{5}$,
Tom Kennedy$^{1}$,
David Linder$^{1}$,
Chris Miller$^{4}$,
Mat Page$^{1}$,
\newauthor
Chris Pattison$^{6}$,
Richard Robinson$^{6}$,
J\"urgen Schmoll$^{2}$,
Dave Walton$^{1}$
and Berend Winter$^{1}$
\\
$^{1}$Mullard Space Science Laboratory, University College London, Holmbury St Mary, Dorking RH5 6NT, UK\\
$^{2}$Department of Physics, Durham University, South Road, Durham DH1 3LE, UK\\
$^{3}$IRFU, CEA, Universit\'e Paris-Saclay, F-91191 Gif-sur-Yvette, France\\
$^{4}$UK Astronomy Technology Centre, Royal Observatory, Blackford Hill, Edinburgh EH9 3HJ, UK\\
$^{5}$RAL Space, STFC Rutherford Appleton Laboratory, Harwell Campus, Didcot OX11 0QX, UK\\
$^{6}$University of Portsmouth, Mercantile House, Hampshire Terrace, Portsmouth PO1 2EG, UK}

\date{Accepted XXX. Received YYY; in original form 29 September 2026}
\pubyear{2026}

\begin{document}
\label{firstpage}
\pagerange{\pageref{firstpage}--\pageref{lastpage}}
\volume{HWO~special~issue}
\maketitle

\begin{abstract}
The Habitable Worlds Observatory (HWO) will be NASA's next flagship astrophysics mission, designed to search for signs of life on planets around nearby stars and conduct transformational science across astrophysics.
Its High Resolution Imager (\HRI) is not only the observatory's main imaging camera, but also provides grism spectroscopy and sub-microarcsecond astrometry, as well as the observatory's fine guidance, wavefront sensing and pupil imaging.
We present a design concept for \HRI, developed during concurrent design sessions by flowing down science requirements within the envelope of NASA's EAC-1 telescope architecture.
Light from a $3'\times2'$ field is collimated by two mirrors, then divided by a dichroic for simultaneous observations in an ultraviolet--visible arm covering 200--900\,nm and a near-infrared arm covering 0.9--2.2\,\um.
By minimising the number of reflective surfaces upstream of the dichroic, our optical design maximises throughput and stability, while reducing wavefront error.
Readout modes can be chosen in orbit to image the entire field in exposures of at least 10~seconds, or in smaller windows for observations up to 1\,kHz.
Throughout, we emphasise the interplay between subsystems that couple design choices.
\end{abstract}


% ---------------------------------------------------------------------
% !TEX root = ../main.tex
\section{Introduction}
\label{sec:intro}

The Habitable Worlds Observatory (HWO) is the large ultraviolet, optical and infrared space telescope recommended by the most recent US decadal survey \citep{decadal2021}.
It is intended to launch around 2040 and, when it does, it will be the most capable telescope ever flown.
Its defining goal is the search for signs of life on planets orbiting other stars, but HWO also inherits the role that Hubble and JWST have played for the wider astrophysics and cosmology community. 
To meet that broad remit, NASA's 2023--25 Science, Technology and Architecture Review Team (START) assembled 76 science investigations spanning exoplanets, the Solar system, stellar and galactic astrophysics and cosmology \citep{scowen2025,scicases2026}.

\begin{figure}
\centering
\includegraphics[width=\columnwidth]{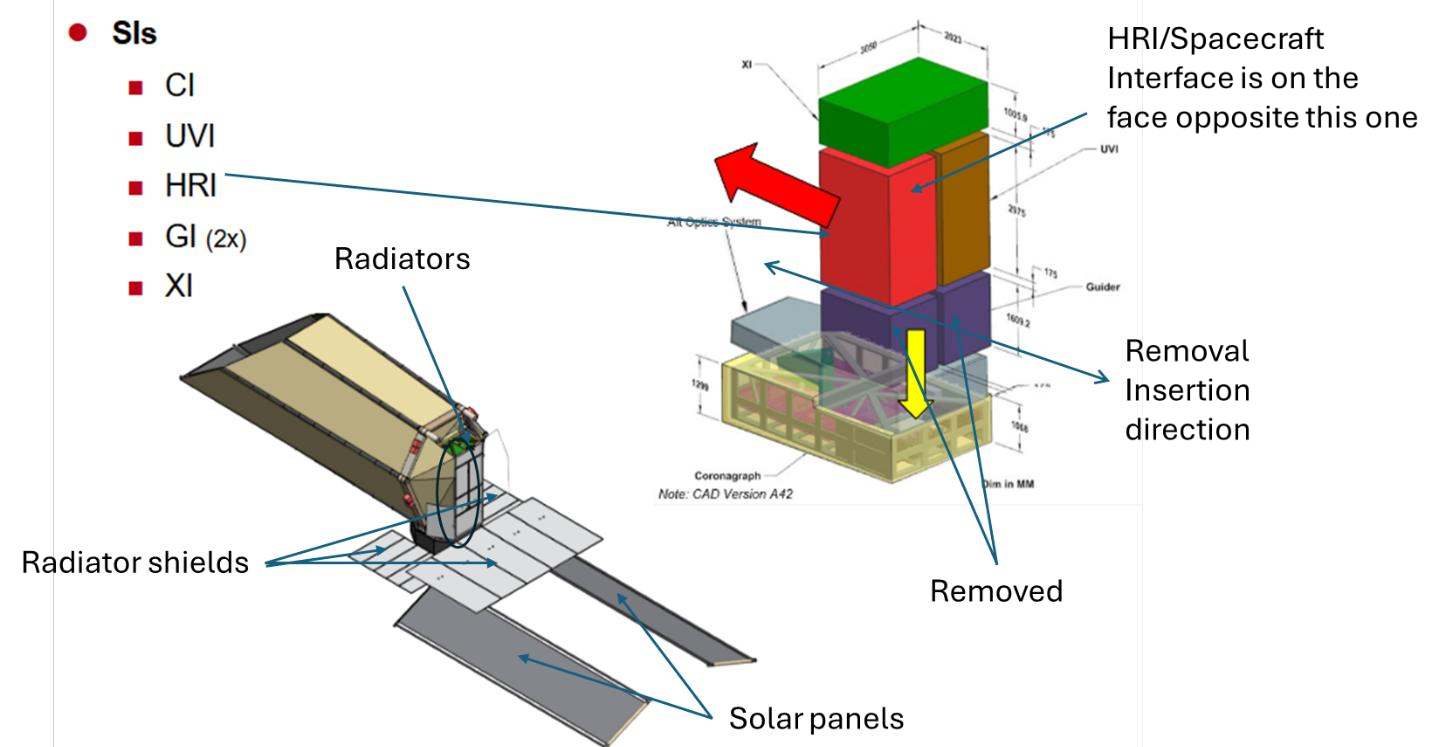}
\caption{\HRI\ in its spacecraft context, based on \citet{liu2026}. The instrument sits behind the primary mirror, with its radiators on the face opposite the primary, viewing cold space; this placement is what makes passive cooling to 75\,K feasible without a mechanical cooler (Section~\ref{sec:thermal}). The astrometric channel shares the same volume allocation, which is why its resources appear throughout the budgets even though its optical design is reported elsewhere.}
\label{fig:context}
\end{figure}

The observatory architecture draws on two earlier large mission concept studies, LUVOIR \citep{luvoir2019} and HabEx \citep{gaudi2020}.
At least three instruments are expected at launch: a coronagraph, a multi-object spectrograph, and a High Resolution Imager (\HRI; see Figure~\ref{fig:context}).
\HRI\ is the observatory's main imaging camera.
Several virtual telescope designs known as Engineering Architecture Concepts (EAC) currently exist \citep{feinberg2024,feinberg2026}. The EAC-1 baseline design calls for imaging over a $3'\times2'$ field with more than fifty filters spanning the ultraviolet to the near-infrared, several grisms for slitless spectroscopy, and an astrometric capability at the sub-microarcsecond level \citep{liu2026}.
It is required by 33 of the 76 investigations, more than any other instrument. A broad summary of the science that motivates the HRI instrument described here, together with the interests of the community that assembled this study, is given by \citet{vaneylen2025}.

Three observatory-level roles add to \HRI's requirements and design.
It should supply telescope fine guidance (from continuous, rapid measurements of astrometry of bright guide stars) to keep the telescope pointed at its science target; wavefront sensing to maintain the phasing and alignment of HWO's segmented primary mirror; and pupil plane imaging to acquire `selfie' images of that mirror, to monitor micrometeorite or other damage.
Together these mean that \HRI\ is not simply the instrument that produces the mission's imaging data, which potentially has the broadest impact; it is a piece of observatory infrastructure.
Unlike other instruments it cannot be deferred to a servicing mission, because without it the telescope cannot be pointed or aligned.

This paper describes a concept study for \HRI\ carried out by a UK-led consortium with a French partner responsible for the astrometric channel.
The study was funded by the UK Space Agency and executed largely through two intensive sessions at a Concurrent Design Facility (CDF), with daily contact with NASA staff so that the design could track the mission architecture as it evolved.
The work was baselined against NASA's EAC-1, a 6\,m off-axis segmented telescope with a room temperature optical train, an $f$-number of 20.4 at the focal plane, and an orbit at the second Earth--Sun Lagrange point L2.
Several other candidate architectures, EACs 2-6, are being considered and the next phase of this study will adopt some of these other designs.

\HRI\ comprises an Imager, itself split into an ultraviolet--visible (UVIS) channel and a near-infrared (NIR) channel, together with an astrometric channel that also supplies the fine guidance function.
The astrometric channel has its own focal plane fed directly by the telescope, with intentionally no foreoptics, to minimise sources of unknown field distortion and to add stability. It is developed by a separate team and is described in detail by \citet{amiaux2025}; its scientific rationale, and in particular the measurement of true masses for the potentially habitable planets that the coronagraph will identify, is set out by \citet{malbet2026}.
We therefore discuss the astrometric channel in this paper only where it shares resources with the Imager.
It sits inside the same enclosure, draws on the same power and data processing units, and is the single largest consumer of the instrument's thermal and power budgets.

We aim to explain the reasoning behind the design rather than to document it exhaustively, because the details will change with further EACs.
Section~\ref{sec:science} traces the flowdown from science cases to numerical requirements.
Section~\ref{sec:optics} describes the optical design, the argument for keeping the number of reflective surfaces small, and the fine guidance, wavefront sensing and pupil imaging functions that the same optics provide for the observatory as a whole.
Section~\ref{sec:detectors} describes the focal planes and, in particular, the readout modes that let one piece of hardware behave as several different instruments.
Sections~\ref{sec:thermal} to \ref{sec:electronics} cover the thermal, mechanical and electronic architectures.
Section~\ref{sec:discussion} draws out the couplings between these subsystems, which we regard as the most important result of the study, and Section~\ref{sec:conclusions} sets out what needs to happen next.

% !TEX root = ../main.tex
\section{Flowdown of science requirements to instrument requirements}
\label{sec:science}

This facility instrument will be used for a broad range of science.
In this section we work through the parameters that drive the design and identify the science that determines each of them.
Throughout, we refer to investigations by the identifiers used in NASA's Science Case Development Documents SCDDs, grouped as Living Worlds (LW), Solar System in Context (SSiC), Evolution of the Elements (EE) and Galaxy Growth (GG).
These were collected by \citet{scicases2026}, and the subset of most relevance to \HRI\ and of interest to the UK community further described by \citet{vaneylen2025}.
Not all science cases have an equally-complete implementation of a Science Traceability Matrix, so we have attempted to follow up with authors of key cases where possible.

\subsection{Driving science cases of \HRI}

%The science drivers of \HRI\ fall into three broad themes.
In exoplanetary science, the coronagraph will identify potentially habitable planets and characterise their atmospheres, but interpreting an atmospheric spectrum requires knowing the planet's surface gravity, and therefore its mass \citep[e.g.,][]{batalha2019}.
HWO relies on \HRI's astrometric channel for those masses \citep{malbet2026}, which places \HRI\ directly on the critical path for the mission's headline result (SCDD-LW-1, SCDD-LW-12).
Separately, HRI's wide wavelength range and grisms support transmission and emission spectroscopy and phase curves of transiting planets from the ultraviolet to the infrared (SCDD-SSiC-16; \citealt{wakeford2025}), and the characterisation of sub-Neptunes and candidate water worlds (SCDD-SSiC-3).

In solar system science, \HRI's ability to track moving targets will also be essential for HWO's role in planetary defence (SCDD-SSiC-30).
\HRI\ is also proposed for observations of ocean worlds including Europa, Enceladus and Ceres (SCDD-SSiC-1; \citealt{cartwright2025}), for Venus as the once and future Earth (SCDD-SSiC-5), and for the giant planets and their ring systems (SCDD-SSiC-7, SCDD-SSiC-29).

Extragalactic astrophysics and cosmology proposed for \HRI\ is exceptionally broad.
Among these, counting substructure in dark matter haloes, through dwarf galaxies (SCDD-GG-12) or gravitational lensing (SCDD-GG-13; Vassilakis et al., this volume), will test whether dark matter is cold and particulate or something else \citep{battaglia2022,vegetti2024,adhikari2025}; the astrometric channel adds a third route through the internal motions of stars in Milky Way satellites, which was the motivation for the Theia concept \citep{boehm2017,malbet2021}.
Resolving individual stars in nearby galaxies will constrain their star formation and chemical evolution (SCDD-EE-6), while Cepheids and the tip of the red giant branch at cosmological distances would remove a rung from the distance ladder (SCDD-EE-10), where the present limiting systematics are set by the calibration of the Cepheid and tip-of-the-red-giant-branch scales \citep{freedman2019,riess2022}. In extragalactic observations of the static sky, knowledge of the PSF is generally key \citep{massey13}, as much as stability.

This list of science cases is non-exhaustive as HRI is envisioned to be a broad-purpose instrument, and \HRI\ will complete many compelling science investigations. These, however, serve as key examples that directly informed some of the instrument design choices.

\subsection{Instrument requirements}

\subsubsection{Field of view}

Almost all of the non-exoplanet programmes benefits from a larger field, and most of them concern the static sky, so a smaller field would merely mean a cost in efficiency.
The specific drivers are the cores of galaxy clusters at moderate redshift, which need $3'\times2'$ to $3'\times3'$ (SCDD-GG-13); the ring systems of Saturn and Jupiter at $2'\times2'$ (SCDD-SSiC-29); protoplanetary discs around nearby stars, also $2'\times2'$ and additionally requiring an occulting mask and Lyot stop of the kind carried by JWST (SCDD-SSiC-8); and Venus and the terrestrial planets at $1'\times1'$ (SCDD-SSiC-5).
We adopted $3'\times2'$, consistent with the NASA baseline.

\subsubsection{Short-wavelength cut-off}

The blue end of the passband is set almost entirely by atomic and molecular species.
Access to hydrogen in Titan's atmosphere requires reaching 120\,nm (SCDD-SSiC-27); carbon monoxide on Mars and elsewhere, 180\,nm (SCDD-SSiC-28, SCDD-SSiC-7); nitrogen, 250\,nm; carbon dioxide, 280\,nm; ozone, 290\,nm; and aerosols in Titan's atmosphere, 330\,nm.
The extragalactic cases push in the same direction for different reasons: a census of supermassive black hole masses and spins wants 100\,nm (SCDD-GG-2), as do studies of the sources that reionised the Universe (SCDD-GG-9, SCDD-GG-11); dust grain properties in the Milky Way and in the tori around active nuclei want 210\,nm (SCDD-GG-3, SCDD-GG-14); and resolving individual stars in Local Group galaxies wants 300\,nm (SCDD-EE-6).

We adopted a cut-off of 200\,nm with a goal of 100\,nm.
The distinction matters for reasons developed in Sections~\ref{sec:optics} and \ref{sec:coatings}: reaching below 200\,nm changes the available coatings, and with them the throughput and the wavefront-sensing error budget.

\subsubsection{Long-wavelength cut-off}

At the red end there is one strong spectroscopic driver, the hydrated minerals on the martian surface that absorb between 1.9 and 2.5\,\um\ (SCDD-SSiC-28), and one strong physical limit.
HWO's mirrors operate at room temperature, so the telescope itself becomes a significant source of thermal background beyond roughly 2\,\um.

The NASA baseline sets a minimum long wavelength cutoff of 1.8\,\um.
We adopted 2.2\,\um\ as a goal for this study and used it as the working baseline, because it captures a scientifically important part of the hydrated-mineral band and the $K$ atmospheric window while staying close enough to the thermal turnover that the background remains manageable.
Where this choice propagates into the design we flag it, because it is one of the more consequential decisions taken: it sets the red edge of every broadband coating, the cut-off of the mercury-cadmium-telluride detector material, and therefore the detector operating temperature and a large fraction of the radiator area.

\subsubsection{Simultaneous coverage, and the case for a dichroic}
\label{sec:simultaneity}

Several cases require observations in the ultraviolet and the infrared at the same time, or within minutes of each other.
This requirement more than any other determines the architecture of the instrument.

Transiting exoplanets require simultaneous imaging and grism spectroscopy from 300\,nm to 1.8\,\um, because a transit happens once and cannot be repeated with a different filter (SCDD-SSiC-16).
Monitoring weather, cloud morphology and surface features on Venus requires simultaneous coverage across the full range (SCDD-SSiC-5), as do the giant planets at a more relaxed cadence (SCDD-SSiC-29).
Removing a rung from the distance ladder by measuring Cepheids and the tip of the red giant branch at cosmological distances requires simultaneous coverage from 500\,nm to 1.8\,\um, so that the colour and magnitude of Cepheid variable stars refer to the same epoch (SCDD-EE-10).

We found that an instrument attempting to reach this entire wavelength range with a single beam cannot satisfy all other requirements.
The alternative is to split the beam to two (or more) focal planes.
The cost is an additional surface and a second set of re-imaging optics, detectors and electronics.
The benefit, developed in Section~\ref{sec:optics}, is that each arm can then be built with the mirror substrate, surface finish and coating appropriate to its own band instead of a compromise across the whole range.
We placed the split at 900\,nm, which suits the short-wavelength limit of the infrared detector material and keeps the silicon channel within the range where its quantum efficiency is useful.
The exact wavelength remains open and will be traded against coating performance and supplier capability.

\subsubsection{Sampling and image quality}

There is little point in building a large, diffraction-limited telescope then undersampling its point spread function.
NASA's requirement specifies diffraction-limited performance at 500\,nm for the UVIS channel and 1000\,nm for the NIR channel.
We designed to a more demanding 420\,nm and 850\,nm respectively, to avoid the instrument limiting observatory resolution if as-built performance exceeds requirements (as it has on most recent space telescopes).
The ratio of these requirements is also compatible with available detector formats, where pixel sizes of 6.5\,\um\ and 10\,\um\ will both sample the diffraction-limited PSF equally well.
At the resulting plate scale a UVIS pixel subtends 9.35\,mas.

\subsubsection{Point spread function stability}

Two science cases set an explicit stability requirement.
Strong gravitational lensing to determine the nature of dark matter (SCDD-GG-13) requires system wavefront error better than 35\,nm rms (Vassilakis et al., this volume). 
Uncertainty in the point spread function propagates directly into the inferred distribution of lens mass, and the ability to identify the substructures that are expected only if dark matter is dynamically cold.
Transit spectrophotometry needs stable detectors and pointing within one thousandth of a pixel to spatially map the atmospheric circulation and chemistry of exoplanets (SCDD-SSiC-16).

The telescope environment is specified to be thermally stable at the millikelvin level, and the instrument design must avoid squandering that.
In practice this drives the isothermal approach to the optical bench described in Section~\ref{sec:mechanical} and the decision to place dissipating elements outside the cold enclosure wherever possible.

\subsection{Baseline instrument requirements}

The instrument requirements adopted for this study are listed in Table~\ref{tab:req}.
Some of these are properly requirements; others such as pixel scale, are design characteristics that could evolve without affecting either performance or the spacecraft interface.

The numbers should be read as design points and ranges rather than guarantees.
A cadence of 10\,Hz does not mean that every science mode needs exactly ten frames each second; it identifies an operating point at which window size, read noise, data rate and spacecraft interface remain in balance.
Likewise the 420\,nm and 850\,nm diffraction limits are targets against which the optical prescription is assessed, not claims that the full observatory will deliver a perfectly sampled Airy pattern at those wavelengths.
The architectural question is not whether each specification could be met on its own, but whether they can be met together without an unacceptable growth in optical surfaces, detector power, mechanism count or data system complexity.
Section~\ref{sec:discussion} returns to this.

To support this work and to allow science cases to be costed in observing time, we wrote an exposure time calculator for \HRI\ in Python, following the approach used for the LUVOIR High Definition Imager \citep{postman_etc}.
It propagates a source spectrum through the optical train, applies filter transmission and detector response, and combines shot noise from the source, from zodiacal light and from dark current with read noise and analogue-to-digital quantisation uncertainty.
Zodiacal light is modelled with \textsc{ZodiPy} \citep{zodipy2024}, whose tabulation begins at 1.25\,\um, extended to shorter wavelengths using the correction derived by the SkySurf programme \citep{windhorst2022,obrien2026}.
Thermal emission from the warm telescope is not yet included, so the calculator is currently most reliable in the UVIS channel.
%It is not yet publicly released.

\begin{table}
\centering
\caption{Baseline requirements adopted for the study. %, following the NASA configuration tracker except where noted.
%The long-wavelength cut-off of 2.2\,\um\ was adopted as a goal for this study and used as the working baseline, enhancing the NASA minimum of 1.8\,\um.
}
\label{tab:req}
\begin{tabular}{lll}
\toprule
Parameter & UVIS & NIR \\
\midrule
Function & imaging, grism, & imaging, grism \\
~ & WFS, guiding & spectroscopy \\
Field of view & $3'\times2'$ & $3'\times2'$ \\
%Bandpass & 200--900\,nm & 0.9--2.2\,\um \\
% & (100\,nm goal) & \\
Short wavelength cutoff & 200\,nm (100\,nm goal) & 0.9\,\um \\
Long wavelength cutoff & 900\,nm & 1.8\,\um\ (2.2\,\um\ goal) \\
Diffraction limit & 420\,nm & 850\,nm \\
Detector format & $8192\times8192$ & $4096\times4096$ \\
Mosaic & $3\times2$ & $3\times2$ \\
Pixel size & 6.5\,\um & 10\,\um \\
Read noise & $<1$\,\ep & $<2.5$\,\ep \\
Dark current & $<10^{-4}$\,\ep/pix/s & 0.002\,\ep/pix/s \\
\midrule
\multicolumn{3}{l}{\textit{Instrument-level requirements}} \\
Mass & \multicolumn{2}{l}{$<450$\,kg (excluding astrometry)} \\
Power & \multicolumn{2}{l}{$<183$\,W (excluding astrometry)} \\
Volume envelope & \multicolumn{2}{l}{$<1.44\times2.02\times2.98$\,m} \\
Radiator area & \multicolumn{2}{l}{$<4.28$\,m$^2$} \\
Interface temperature & \multicolumn{2}{l}{293\,K} \\
\midrule
\multicolumn{3}{l}{\textit{Telescope (EAC-1)}} \\
Aperture & \multicolumn{2}{l}{6\,m inscribed, off-axis, segmented} \\
Focal ratio & \multicolumn{2}{l}{$f$/20.4} \\
Diffraction limit & \multicolumn{2}{l}{500\,nm} \\
Orbit & \multicolumn{2}{l}{L2 quasi-halo} \\
\bottomrule
\end{tabular}
\end{table}

% !TEX root = ../main.tex
\section{Optical design}
\label{sec:optics}

\subsection{Foreoptics}
\label{sec:collimator}

\begin{figure*}
\centering
\includegraphics[width=0.92\textwidth]{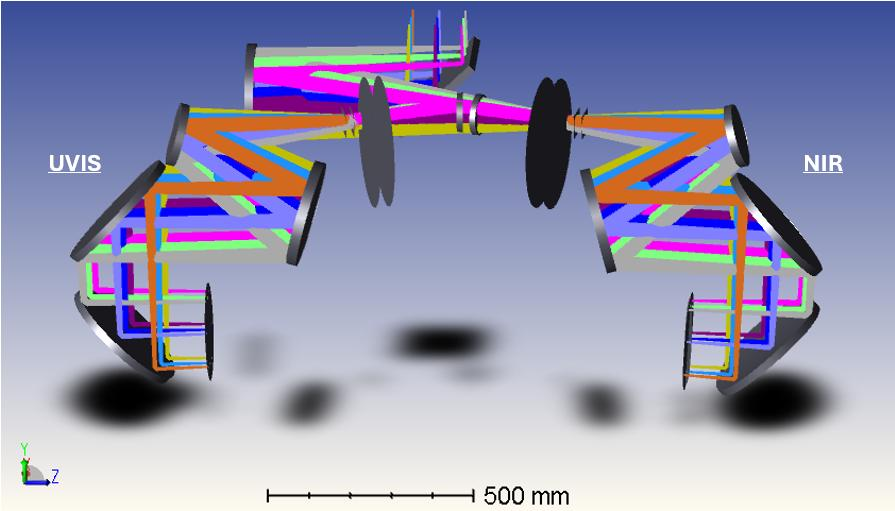}
\caption{Optical layout of the \HRI\ Imager.
Light from the telescope focal plane is folded by COL-M1, collimated by COL-M2 to a 40\,mm pupil where the forward interchange mechanism divides the beam, and re-imaged by four surfaces in each arm.
Seven reflections separate the telescope from the UVIS detector and six from the NIR detector, which is the shortest path we found that meets the requirements of Section~\ref{sec:science}, and the basis of the throughput and coating arguments of Section~\ref{sec:coatings}.}
\label{fig:layout}
\end{figure*}

As outlined in Section~\ref{sec:intro}, the Imager splits into a UVIS and NIR channel. We use only two mirrors ahead of this split (see Figure~\ref{fig:layout}), to minimise the number of surfaces that need a broadband coating reaching from 200\,nm to 2.2\,\um.
NASA's baseline \HRI\ concept in EAC-1 uses four mirrors.
Of our two, COL-M1 is a flat fold mirror, necessary to avoid any hardware ending up in front of the HWO focal plane. The single powered surface COL-M2 collimates the beam, avoiding the introduction of astigmatism by the tilted dichroic beam splitter for the transmitted NIR channel. While the collimation is sufficient to avoid the astigmatism, it does not produce a diffraction limited flat wavefront. The residual aberrations are corrected for within the two imager trains instead. 

Light enters the instrument from the EAC-1 telescope focal plane at a large tilt, and the focal plane is not telecentric.
To invert this, COL-M2 receives the field at $12^{\circ}$ angle of incidence, and the same steep geometry propagates into the arms, where M1 and M2 of both channels also operate near $12^{\circ}$.
Angles this far from normal produce strongly anamorphic footprints, which the downstream surfaces correct with high local curvature and freeform figures of high Zernike order.
The tilted input is therefore the origin of both the unusual incidence angles quoted throughout this section and the manufacturing difficulty discussed in Section~\ref{sec:optics-performance}.

COL-M1 is a flat fold at $45^{\circ}$, $180\times260$\,mm.
COL-M2 is a toroidal mirror, nearly spherical, 250\,mm in diameter, which collimates the beam and provides a pupil of 40\,mm diameter.
That pupil size allows transmission through a set of narrow filter wheels holding filters of 50\,mm diameter, which keeps the wheels, bearings and motors small enough to sit inside the cold enclosure without dominating either the mass budget or the dissipation within it.
Locating the filters at an intermediate pupil minimizes their diameter, while local transmission variations across any filter are averaged out over the whole field of view. 
Both collimator mirrors are 30\,mm thick Ultra-Low Expansion (ULE) Corning glass, ground and polished, specified to form error $<$50\,nm peak-to-valley and roughness $<$1\,nm rms in order to remain diffraction-limited at 420\,nm.

\subsection{Forward interchange mechanism}
\label{sec:fim}

\begin{figure}
\centering
\includegraphics[width=\columnwidth]{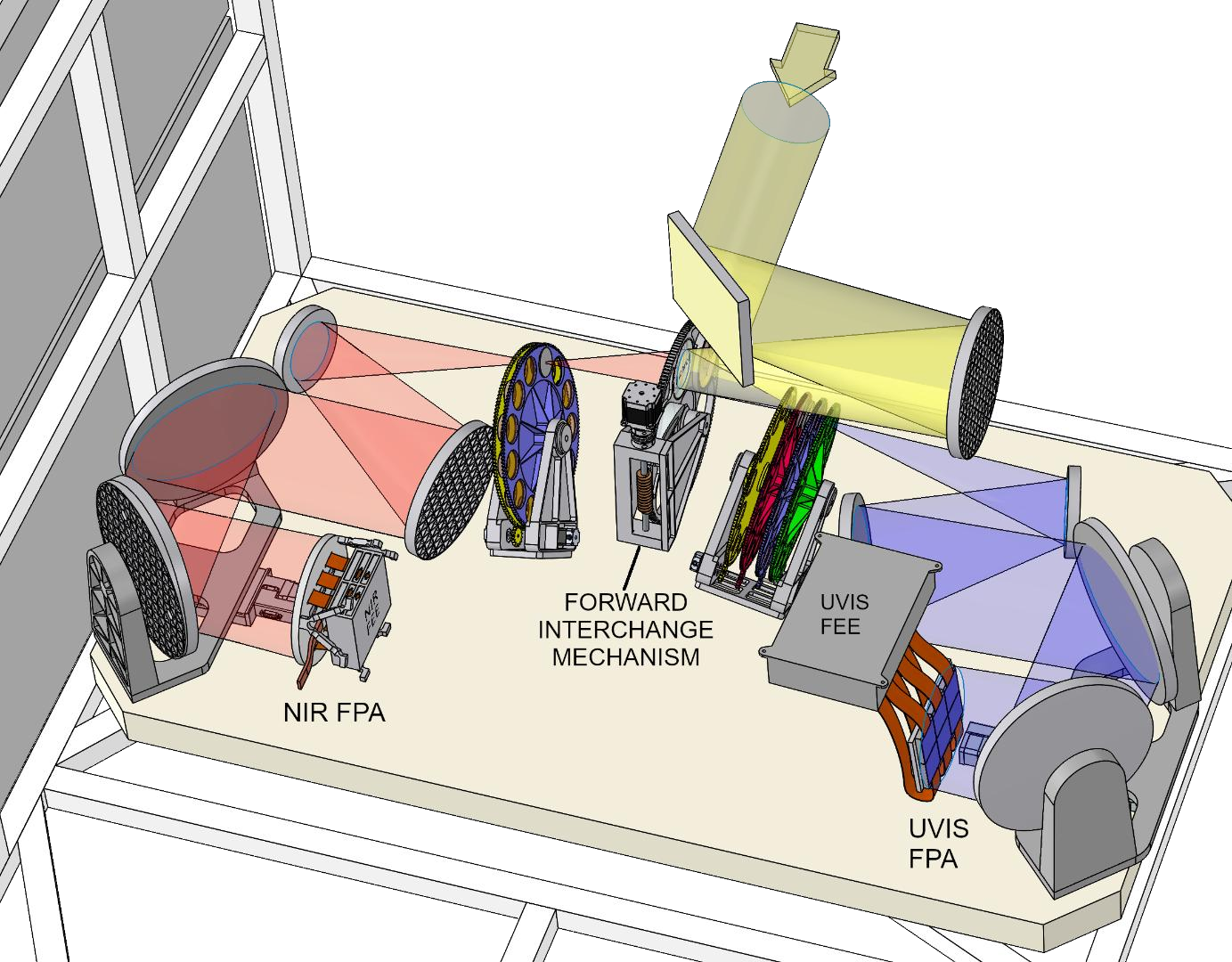}
\caption{The optical path as accommodated, with the UVIS arm to the right and the NIR arm to the left, showing that the layout of Figure~\ref{fig:layout} packages into the required volume without contortion.
The forward interchange mechanism sits at the centre, ahead of the split, so that one mechanism reconfigures both arms.
The UVIS front-end electronics are immediately behind the UVIS focal plane assembly but outside the cold enclosure, which keeps the largest dissipator in the imager out of the volume that must reach 178\,K.}
\label{fig:cadpath}
\end{figure}

The optical path then encounters the forward interchange mechanism (FIM), a wheel with six positions that reconfigures the instrument at the level of its architecture rather than its passband (see also Figure~\ref{fig:cadpath}:

\begin{enumerate}
\item a 900\,nm cut-off dichroic, for simultaneous two-arm operation.
This is a 100\,mm fused silica flat used at $12^{\circ}$, reflecting the UVIS arm and transmitting the NIR arm, with fused silica chosen for its transmission to 2.2\,\um.
The transmitted beam suffers a small chromatic aberration, so a plate of the same thickness and opposite tilt is added to compensate; the residual shift is under five per cent of the Airy disc and this corrector may prove unnecessary.
\item a flat mirror, for UVIS-only measurements at higher efficiency or over an enlarged wavelength range. This means that the dichroic element can be replaced by a mirror of higher efficiency for any observation that does not need simultaneity.
\item an empty aperture, providing the same function mentioned in (ii) for the NIR channel.
\item scatter plates for UVIS and NIR flat-field calibration, back illuminated by the calibration units.
\item a concave mirror of 1000\,mm radius of curvature, imaging the entrance pupil onto the UVIS detector.
\item a spare position for an alternative dichroic, optimised for short or long wavelength performance, or a 50/50 beam splitter.
\end{enumerate}

\begin{figure*}
\centering
\includegraphics[width=0.95\textwidth]{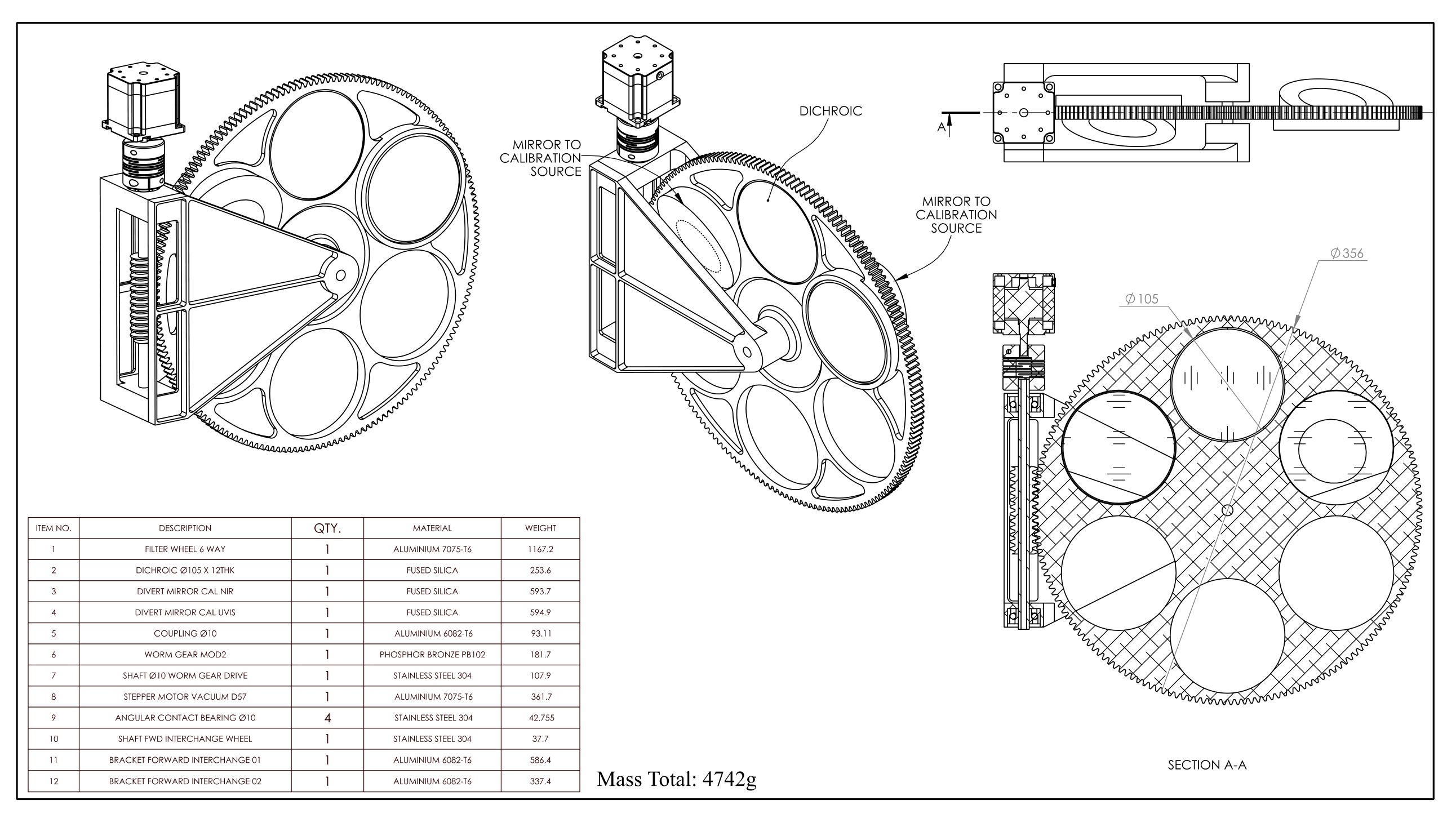}
\caption{The forward interchange mechanism.
Six positions reconfigure the instrument between simultaneous two-arm operation, single-arm operation at higher throughput, flat-field calibration and pupil imaging, so that capabilities usually requiring separate optical paths are obtained from one 4.7\,kg wheel.
Two positions carry mirrors that divert light from the calibration sources.
The worm drive at upper left provides the repeatable angular positioning whose required accuracy is discussed in Section~\ref{sec:wfs}.}
\label{fig:fim}
\end{figure*}

A worm drive is used to position the FIM, giving fine and repeatable rotation at the price of some complexity; a pinion drive would be simpler and may prove sufficient once positioning requirements are set.
A comparable worm-driven wheel was developed for the XIPE mission concept \citep{zane2016}.
Figure~\ref{fig:fim} shows the assembly, with estimated mass 4.7\,kg.

Note that we began the design against the EAC-1 telescope from first principles, without reference to any previous solution.
Nonetheless, we arrived at an architecture closely resembling the High Definition Imager of the LUVOIR study \citep{luvoir2019}: a short common collimator, an intermediate pupil carrying the spectral selection hardware, and a dichroic split into separate ultraviolet-visible and near-infrared re-imaging trains.
That two independent routes reach the same place suggests the architecture is close to forced by the requirements.
Our FIM, however, adds the calibration and pupil-imaging positions listed above, so that one mechanism serves several observing and calibration roles that would otherwise need dedicated hardware.

\subsection{The ultraviolet--visible arm}

Achieving the diffraction limit at 420\,nm imposes the same surface requirements on the UVIS arm as on the common foreoptics: form error of 50\,nm peak-to-valley and roughness of 1\,nm rms.
The mirrors must therefore be manufactured by CNC grinding and polishing of ULE glass substrates.
They are currently modelled as solid discs 20\,mm thick and would be lightweighted in a later iteration.
The mirror coatings no longer need a coating that is reflective to 2.2\,\um,

Four surfaces re-image the field onto the detector mosaic.
UVIS-M1 is a Zernike freeform 160\,mm across, described by 24 coefficients from $Z_4$ to $Z_{27}$, used at $12^{\circ}$.
UVIS-M2 is a biconic concave mirror of 270\,mm, also at $12^{\circ}$.
UVIS-M3 is a milder freeform of 330\,mm diameter using seven coefficients from $Z_4$ to $Z_{10}$, at $45^{\circ}$, and UVIS-M4 is a concave cylinder of 310\,mm, again at $45^{\circ}$.
A focus mechanism is required, implemented as a pair of mirrors translating together to adjust path length independently of the other arm.

\subsection{The near-infrared arm}

The NIR arm needs only to be diffraction limited at 850\,nm, which loosens its requirements to form error $<100$\,nm peak-to-valley and roughness $<5$\,nm rms.
Such mirrors can be manufactured by single-point diamond turning of gold-coated aluminium 6061, which is faster, cheaper, straightforwardly lightweighted from the rear, and already flight proven at TRL 9 through JWST/MIRI \citep{wright2023}.
The 27th-order freeform is not a problem for diamond turning, and even higher orders would also be possible.

The point of the dichroic is so that each channel can be manufactured for the shortest wavelength at which it must be diffraction limited, but no more.
%The instrument does not try to make every subsystem optimal to the same standard.
A single-arm instrument covering the full wavelength range would need every surface to satisfy both sets of requirements. 
Its coating would be suboptimal for the narrower wavelength ranges, and it would be considerably more expensive.

\subsection{Pupil plane `selfie' imaging of the primary mirror}
\label{sec:pupil}

\begin{figure}
\centering
\includegraphics[width=\columnwidth]{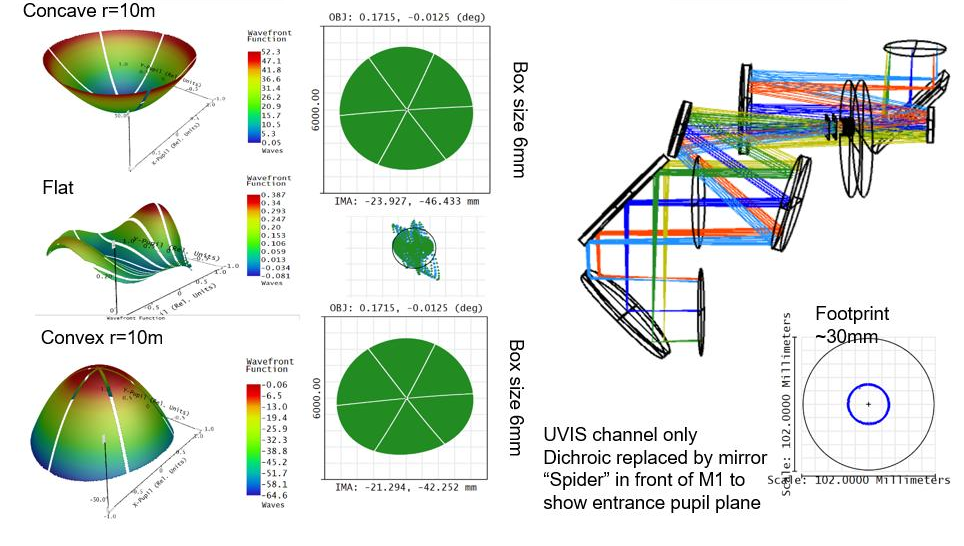}
\caption{Modelling of the pupil imaging function.
Inserting a concave mirror of 1000\,mm radius at the forward interchange mechanism places on the UVIS detector a resolved image of the telescope entrance pupil, and therefore of the segmented primary.
The segment pattern is recovered with enough fidelity to directly diagnose localised contamination and segment-level anomalies.
The direct pupil measurement is complementary to focal-plane wavefront sensing because some micrometeoroid-induced features can be visible in pupil images without a commensurate wavefront-error signal \citep{mcelwain2023}.}
\label{fig:pupil}
\end{figure}

The forward interchange mechanism includes one slot with a concave mirror.
Together with another optical element, which folds into the downstream optical path (but springs out if power is lost), this images the entrance pupil onto the UVIS detector, producing a direct image of the segmented primary mirror  (Figure~\ref{fig:pupil}).
The function is available on the UVIS channel only, where the spatial resolution is highest.

This capability was added during the study rather than being imposed as a requirement.
%It is cheap: one mirror in the wheel of the forward interchange mechanism, which has to exist anyway.
Pupil images in JWST/NIRCam have become part of routine monitoring, after revealing micrometeoroid impacts onto the primary surface %that did not always produce a corresponding change in the measured wavefront error 
\citep{mcelwain2023}.

\subsection{Fine guidance}
\label{sec:guidance}

A second observatory-level function provided by \HRI\ is fine guidance from bright star astrometry. 
This is principally provided by the astrometric channel, which has the largest field of view, no filters (or any optics) and hence access to bright stars.
However, redundancy is provided by the UVIS channel, and probably the NIR channel, for when the astrometric channel is itself acquiring science data and also to militate against risk of failure of the astrometric channel.
If the arm is dedicated to guiding, it would be possible to use no filter, or an extra-wide band filter to truncate red leak that could broaden the PSF.
If used for guiding during a science exposure, all the detectors can be windowed for fast readout of a guide star without disrupting other pixels (see Section~\ref{sec:modes}).
We assume a baseline sampling rate of 10\,Hz inside the window, similar to Roman \citep{stoneking2025}.
Moving targets for planetary defence and Solar system science is accommodated through the moving-window mode of the astrometric channel and through mode~2 of the imager (Table~\ref{tab:modes}).
Our current design does not yet accommodate dedicated data processing with guaranteed latency and update rates, but it maintains logical separation from and well-defined internal interfaces with the science data CDPU, so this can be added later without redesign.

Roll control requires more than one guide star, a point raised by this study with NASA's HWO Technology Maturation Project Office.
JWST's Fine Guidance Sensor closes the fine-guidance loop in pitch and yaw, while roll about the line of sight is controlled using the star trackers and inertial rate units \citep{vila2025}.
Guiding with two stars on a single \HRI\ channel is possible but gives only a short lever arm for roll; guiding simultaneously on two channels would separate the stars further on the sky and constrain roll correspondingly better.
The operational choice will propagate into the limiting magnitude required of a guide star and hence into the fraction of sky on which the observatory can point, although with a 6--8\,m primary aperture we do not anticipate difficulty finding suitable stars anywhere on the sky.

Requirements from NASA are awaited on the number of guide stars, the form in which their positions are reported, and the handling of moving-object guidance for Solar system science and planetary defence.
Flight experience from JWST suggests that guide-star catalogues, acquisition logic, bad-pixel maps and recovery from failed acquisitions should be treated as observatory-level requirements from the start.
JWST's current success rate for identifying and guiding on the commanded star is about 96 per cent, with failures caused by catalogue errors, bad pixels or source misclassification \citep{vila2025}.
Nonetheless, JWST's Fine Guidance Sensor supplies guide-star centroids every 64\,ms, with typical positional accuracy of about 1\,mas, and the normal flight system achieves a typical pointing stability of about 1\,mas \citep{mcelwain2023,vila2025}.

\subsection{Wavefront sensing to align primary mirror segments}
\label{sec:wfs}

\HRI\ will also provide data from which piston and higher-order aberrations are retrieved for each mirror segment. 
These can be adjusted on orbit, to maintain the figure and phasing of the primary mirror.
Wavefront sensing is in addition to pupil imaging, which provides an independent view of the segments themselves (Section~\ref{sec:pupil}).

Building upon the wavefront-sensing approach in JWST NIRCam, the UVIS filter wheels include two dispersive elements and three weak lenses \citep[$\pm8$\,waves of defocus mounted in one filter wheel, and $+4$\,waves of defocus in another;][]{acton2012}.
Combinations of these lenses provide imaging at six focus positions including nominal. 
The same configuration in JWST NIRCam has been used approximately every two days for focus-diverse phase retrieval, and on longer timescales as the measured wavefront error evolved \citep{rieke2023,mcelwain2023}.
In \HRI, wavefront sensing will be confined to the UVIS arm, where the plate scale is finest and the point spread function best sampled, although it affects the choice of UVIS mirror coatings (Section~\ref{sec:coatings}).
Associated data processing is not yet defined, and may reside onboard the instrument or on the ground.

Wavefront sensing will be performed through a $\sim$1~per cent narrowband filter, to keep the point spread function effectively monochromatic, so that phase retrieval carries no broadband smearing term.
We baseline F502N in HST/WFC3, which has 6.5\,nm width at pivot wavelength 501\,nm \citep{wfc3ihb}.
A band near 500\,nm matches the telescope's nominal diffraction limit, and working near the middle of the UVIS coating band ensures that the reflection phase varies least rapidly with wavelength (see Section~\ref{sec:coatings}).
This is especially important for \HRI\ because the local angle of incidence varies across a freeform surface. 
The mirror coating already contributes a phase that depends on pupil position; if that phase also changes appreciably across the filter, the pupil map will differ from one end of the band to the other and a single monochromatic forward model no longer describes it.
Mid-band the map is effectively fixed, leaving a static term that can be calibrated rather than a chromatic one that must be integrated over.
However, note that F502N is centred on the [O\,\textsc{iii}]\,5007 emission line, and nebular emission could elevate the background noise level in some star-forming and crowded fields.
This can be avoided by using a continuum band of the same width placed just off the line. 
To save one spot in a filter wheel, in JWST/NIRCam the equivalent filter is attached to the $+4$\,wave defocus lens \citep{acton2012}.

\subsection{Mirror coatings}
\label{sec:coatings}

NASA is developing protected aluminium coatings for HWO mirrors M1--M4, specifically focusing on enhanced lithium fluoride (eLiF) and xenon difluoride-passivated lithium fluoride (XeLiF) \citep{tuttle2024}.
Replicating this would be an obvious choice for \HRI\ foreoptics mirrors COL-M1 and COL-M2 that require reflectivity over the full wavelength range.
In the absence of this, our design principle would be to prioritise reflectivity at ultraviolet and optical wavelengths, because that offers the best absolute spatial resolution (SCDD-GG-12,13, SCDD-EE-6,10, SCDD-SSiC-16) and is where HWO offers a capability that no other facility of its era will have.
The science cases that require a 200\,nm short wavelength cut-off (SCDD-GG-2,3,9,11,14, SCDD-EE-6, SCDD-SSiC-7,28) cannot use Roman or JWST, which are insensitive below 480\,nm and 600\,nm respectively.
Photons lost to coatings at the blue end of the band are therefore worth more than photons lost at the red end.

Mirrors in the near-infrared arm will be coated with gold, as already flight proven at TRL 9 through JWST/MIRI \citep{wright2023}. 

Coatings for mirrors in \HRI's ultraviolet--visible arm have not yet been selected. 
Over the shorter wavelength range, dielectric multilayer coatings can provide higher reflectivity than metallic coatings, but introduce a more structured phase and polarisation response in the reflected wave.
It remains to be investigated whether this degrades wavefront phase-retrieval methods, which measure the optical path through the telescope.
No problem arises from a spatially constant phase shift, which is absorbed into the choice of reference phase or focus.
However, wavefront aberrations in the reflected beam can arise through spatial variation in the phase shift that depends on angle of incidence, polarisation or wavelength.
Even if coating-induced phase errors are only a few milliwaves near the coating's central wavelength, they are enhanced by fast, broadband or high-incidence-angle systems \citep{reiley1994,liang2001}.

Reflection angles for UVIS-M1 and UVIS-M2 are near $12^{\circ}$, the local angle of incidence varies across a freeform surface, and light arriving from different segments of the primary can illuminate different parts of each \HRI\ mirror.
A coating phase that depends on angle can therefore become a field- and segment-dependent phase term.
In a single-wavelength focal-plane measurement, the lowest-order part of such a term can appear as spurious segment piston.
It is not necessarily indistinguishable from a real piston if multiple wavelengths, field positions or polarisation are considered, but an optical model that neglects the coating can bias the retrieved segment errors.

The next phase of this study can make this quantitative without requiring a new optical design method.
For each candidate coating stack, a thin-film calculation can first provide the complex reflection coefficients $r_s(\lambda,\theta)$ and $r_p(\lambda,\theta)$ over the \HRI\ wavelength range and the full range of incidence angles encountered on each mirror.
These coefficients can then be applied in a polarisation-aware ray trace of the as-designed freeform surfaces.
For each primary-mirror segment, the resulting complex phase can be sampled over the illuminated footprint and fitted to piston, tip, tilt and higher-order terms.
The quantity of most direct relevance to segment phasing is the differential piston between segments after removing common and low-order terms, evaluated for the actual filters and wavefront-sensing bands.
Repeating the calculation over field position and wavelength would show whether the effect is common-mode, a calibratable field term, or a bias that aliases into the segment errors being recovered.
The same model can also propagate coating-induced polarisation and amplitude changes into the point spread function.
Ground measurements of the coated optics, together with polarisation-sensitive and wavelength-dependent modelling, should establish whether the effect is negligible or needs to enter the wavefront-sensing error budget.
Measured witness samples of the flight coating could then replace the nominal layer thicknesses and optical constants in the model, providing a route from coating metrology to the wavefront error budget.
%The relevant quantity is the spatially varying part of the phase after the actual coating prescription and incidence-angle distribution have been propagated through the optical model.

Coating qualification also scales with surface count.
Coatings that perform to 200\,nm, let alone to the 100\,nm goal, are not commodity parts, and each broadband surface is a separate development item that must be specified, procured, witnessed and life-tested.
Contamination accumulates per surface too, and matters disproportionately in the near-ultraviolet where a small loss of transmission is expensive; the cleanliness requirements derived in this study are no more severe than those already met by Euclid, Ariel and PLATO.

\subsection{Filter wheels}
\label{sec:filters}

The science programme calls for a large filter complement, and our study produced a provisional list of 40 filters and grisms in the UVIS arm plus 18 in the NIR arm that should meet most observing goals (Table~\ref{tab:filters}).
Where possible, we specified precise bandpasses by modelling the UVIS filters on those in HST/WFC3 (which mimic Sloan and tile cleanly in wavelength) or HST/ACS (which overlap for higher throughput and better photometric redshifts) and the NIR filters on those in JWST/NIRCam.
Following the successful launch of Roman, we would consider adding F106, F129, F158, F184, F213, which cover a slightly broader wavelength range \citep{akeson19}.
Inheriting passbands from previous missions would simplify archival comparison, and means that photometric calibration strategies have precedent.
However, further iteration with science teams would usefully hone the filter selection.

\begin{table*}
\centering
\caption{The provisional filter complement: how many elements of each type, and which passbands they are.
%UVIS names follow HST/WFC3 and NIR names JWST/NIRCam wherever an equivalent passband exists, so that archival comparison is straightforward and photometric calibration strategies have precedent; grisms, neutral-density filters and wavefront-sensing elements are named by function instead.
Quoted wavelengths are indicative rather than specifications.
%Separating the filters and grisms from the wavefront-sensing elements reconciles the two UVIS totals: 35 elements serve observations and five more serve the telescope, and the 40 together leave eight of the 48 available positions spare for growth.
The three weak lenses sit at $+8$, $-8$ and $+4$ waves of defocus, the last carrying the narrowband coating discussed in Section~\ref{sec:wfs}, exactly as JWST/NIRCam pairs its $+4$ wave lens with a narrowband filter.
%NIRCam carries a single extra-wide filter in each of its channels, and the short-wavelength member, F150W2, has half-power points at 1.007 and 2.38\,\um\ \citep{jdox_nircam_filters}, so one such element covers the whole \HRI\ near-infrared arm at the highest throughput the channel can offer.
%The ultraviolet--visible element is cut at 800\,nm rather than run to the detector limit, so that its red edge lies well inside the dichroic's reflection band.
%A guide star observed through it is then measured in a passband set by the filter, rather than one set by the dichroic's as-built transition and the silicon response beyond it, which would otherwise vary with the colour of the star.
}
\label{tab:filters}
\small
\begin{tabular}{llp{11.4cm}}
\toprule
Element type & $N$ & Elements \\
\midrule
\multicolumn{3}{l}{\textit{Ultraviolet--visible arm: four wheels of twelve elements, 48 positions}} \\
Broadband & 7 & F225W, F275W, F336W, F475W ($g'$), F606W, F625W ($r'$), F775W ($i'$), F814W, F850LP ($z'$) \\
Extra-wide & 2 & 350--900\,nm (like F350LP), 200--900\,nm (like F200LP) \\
Medium band & 8 & F390M, F410M, F467M, F547M, F621M, F689M, F763M, F845M \\
Narrow band & 12 & F280N, F343N, F373N, F395N, F469N, F487N, F502N, F631N, F645N, F656N, F658N, F673N \\
Grisms & 3 & full band (200--900\,nm), ultraviolet (200--400\,nm), optical (390--900\,nm) \\
Neutral density & 3 & 2.5, 5.0 and 7.5\,mag \\
Wavefront sensing & 5 & three weak lenses and two dispersive elements \\
\addlinespace
\textit{All elements} & \textit{40} & eight positions spare \\
\midrule
\multicolumn{3}{l}{\textit{Near-infrared arm: two wheels of twelve elements, 24 positions}} \\
Broadband & 3 & F115W ($J$), F150W ($H$), F200W ($K$) \\
Extra-wide & 1 & F150W2 (1.0--2.4\,\um) \\
Medium band & 5 & F098M, F140M, F162M, F182M, F210M \\
Narrow band & 4 & F953N, F164N, F187N, F212N \\
Grisms & 2 & full band (0.9--2.2\,\um), short band (0.9--1.7\,\um) \\
Neutral density & 3 & 2.5, 5.0 and 7.5\,mag \\
\addlinespace
\textit{All elements} & \textit{18} & six positions spare \\
\bottomrule
\end{tabular}
\end{table*}

\begin{figure*}
\centering
\includegraphics[width=0.95\textwidth]{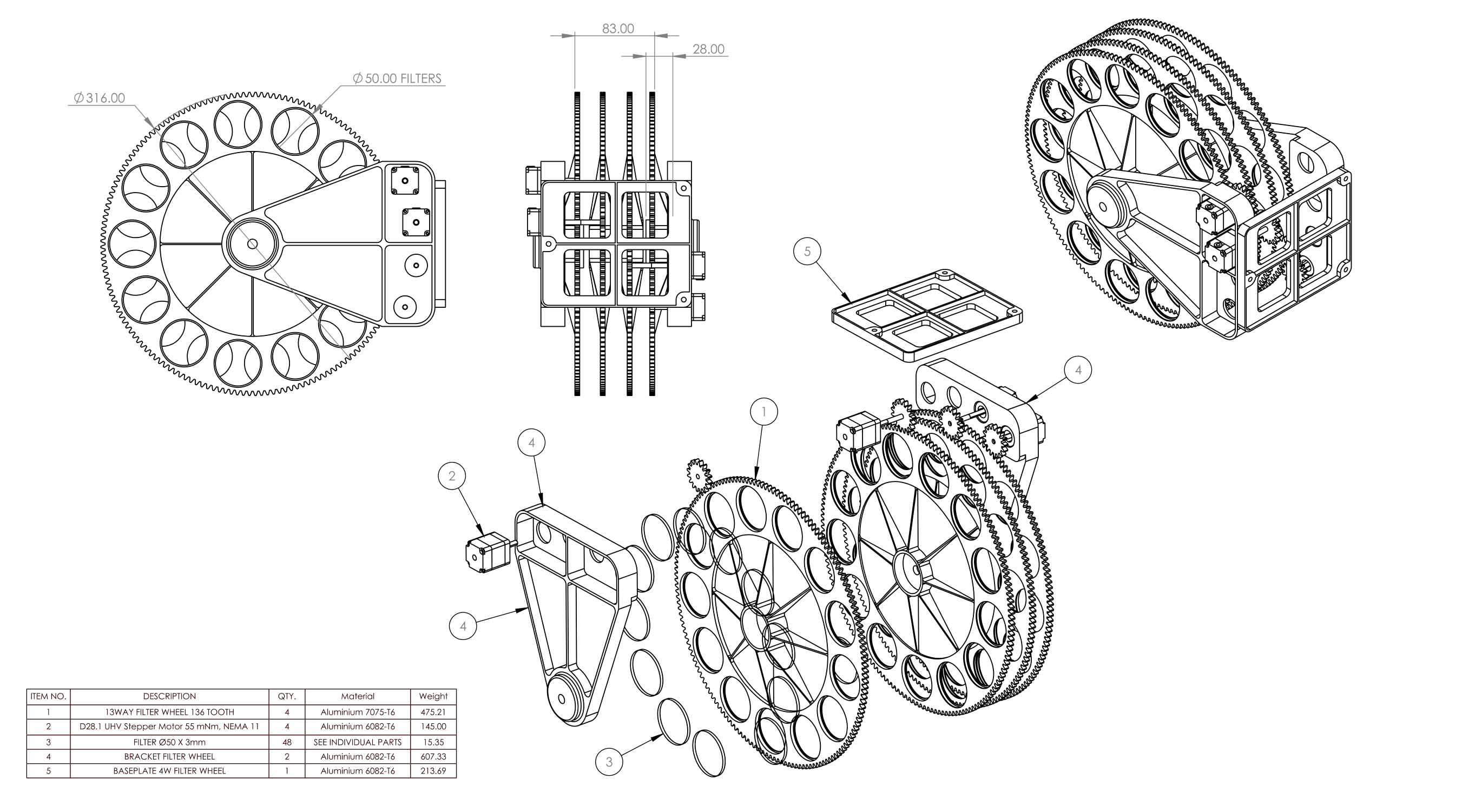}
\caption{Filter wheel concept for the UVIS and NIR arms. Wheels are stacked in pairs on a common axle, each carrying twelve 50\,mm elements and a through hole. This gives access to up to 48 passbands when presenting one element at a time to the beam, and keeps the swept volume small enough to sit inside the cold enclosure. Driving the pinions from opposite sides balances the load on the shared axle.}
\label{fig:filterwheel}
\end{figure*}

In our design, each wheel holds twelve elements plus a through hole to the next, so a stack of four wheels gives access to up to 48 optical elements when presenting one at a time to the beam (see Figure~\ref{fig:filterwheel}).
The wheels are driven by a pinion on the outer circumference, and where two pairs share a common axle, four ultra-high-vacuum stepper motors with space heritage drive them from opposite sides.
Bearings are hybrid silicon nitride, dry lubricated with tungsten disulphide to operate colder than 80\,K in vacuum.
The bearings supporting each wheel are spaced as far apart as the envelope allows, for stability as much as for friction.
Each wheel assembly is self-contained and can be qualified, built and acceptance tested away from the instrument.

We have not yet specified a method for sensing the position of the wheels.
Viable solutions include Hall-effect sensors, photodiodes illuminated through holes in the wheel, or quadrant photodiodes for extreme accuracy.
The choice depends on a repeatability requirement that does not yet exist, but is likely to come from wavefront sensing rather than imaging (see Section~\ref{sec:wfs}).

\subsection{Current performance, and remaining optimisation}
\label{sec:optics-performance}

\begin{figure*}
\centering
\includegraphics[width=0.9\textwidth]{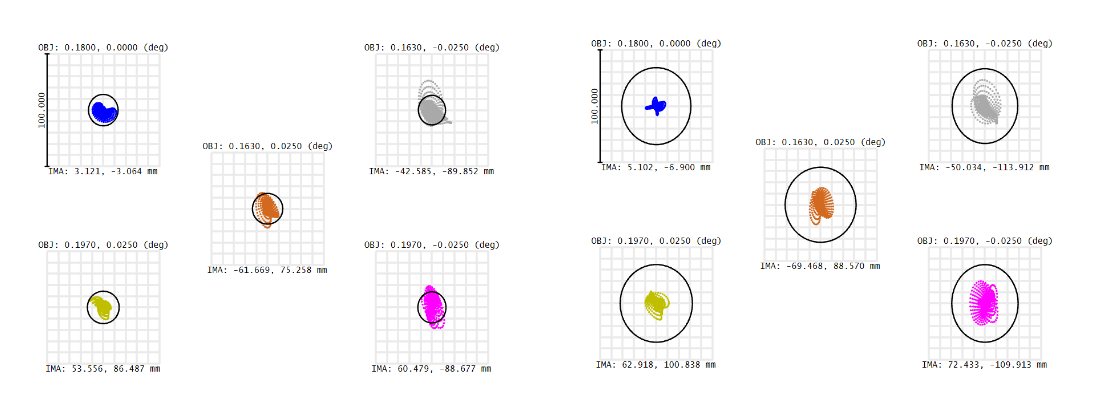}
\caption{Spot diagrams at the five field points used in the initial optimisation, the field centre (blue) and four points at the field edge, with circles marking the Airy disc at 420\,nm (UVIS, left) and 850\,nm (NIR, right).
The NIR spots lie clearly within the disc and the UVIS spots nearly inside it, so the architecture reaches the diffraction limit in the infrared and comes close to it in the ultraviolet-visible using only two mirrors in the common path.
This is the central feasibility result of the study.}
\label{fig:spots5}
\end{figure*}

\begin{figure*}
\centering
\includegraphics[width=0.9\textwidth]{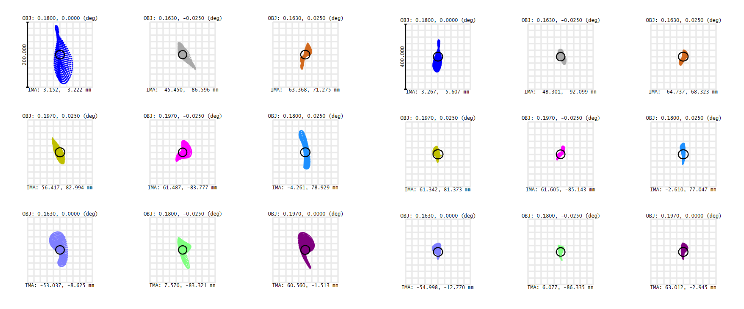}
\caption{The same design re-optimised with four additional field points between the corners.
Image quality at the newly sampled positions falls outside the diffraction limit, showing that the performance of Figure~\ref{fig:spots5} is not yet uniform across the field.
The cause is the high spatial-frequency curvature of the freeform surfaces combined with beam footprints that barely overlap between field points; the likely remedy is one or two further surfaces, at a cost in throughput.}
\label{fig:spots9}
\end{figure*}

Quality analysis of the initial design looks promising.
Mirror surfaces were optimised using raytraced spots at five field points. 
These are well within the 850\,nm Airy disc in the NIR arm, and almost inside the 420\,nm disc in the UVIS arm (Figure~\ref{fig:spots5}).
Modelled optical transmission is 83 per cent.
Grid distortion varies between 7.6 per cent in the NIR arm and 7.1 per cent in the UVIS arm. 
A distortion of this size is not a problem if it is stable and can be measured from dense astrometric fields, but we intend to further study its interaction with variation of the point spread function and with gaps in the detector mosaic, since the mapping from sky to pixel must remain stable enough to register a source between filters and between epochs without calibration becoming the dominant error.

However, the optical design is not yet diffraction-limited everywhere in the field of view.
Adding four field points between the corners and re-optimising the surfaces shows performance outside the diffraction limit at the newly sampled positions (Figure~\ref{fig:spots9}).
This is because the system is fast, so the beam footprint from any one field point covers only a small part of each mirror. The optimiser is therefore free to add small-scale curvature and sculpt each surface locally to favour whichever points it has been given, at the expense of everything between them.
The departure of UVIS-M1 from its best fit sphere shows this directly: an asymmetric form with structure on scales similar to the separation between footprints.
Comparable field dependence exists to some degree in NASA's baseline design, and the effect is a property of fast freeform re-imagers rather than of this particular solution.
The current point design should therefore be read conservatively. 
It demonstrates that the architecture closes, but leaves its optimisation to a phase in which the telescope model, and therefore the input beam, are stable.

Three improvements are possible. 
Slowing the optical system and separating the surfaces so that footprints overlap more would reduce their spatial frequency content and hence the manufacturing and metrology cost, at the price of volume.
Denser field sampling during optimisation is necessary but may be insufficient on its own, since it constrains the surfaces without giving the design the freedom to satisfy the constraints.
Adding one or two surfaces would supply that freedom and is the most likely solution.
The design has substantial margin in both volume and mass, but each added surface reduces the throughput and tightens requirements on mirror coating (see Section~\ref{sec:coatings}).
Note that, for a diffraction-limited instrument, a small low-order wavefront error changes the encircled energy considerably more than it changes geometric spot size.
Final optimisation should therefore minimise wavefront error rather than raytraced quantities.
Requirements on mirror figure, decentre, tilt, focal plane position, detector flatness, filter wedge and thermal sensitivity should be expressed in budgets of wavefront and encircled energy.

The next phase of this study should also perform a stray light analysis. 
The dichroic, filter wheels and mechanism cavities all offer opportunities to generate an unwanted reflection when a bright source lies outside the science band.
The figure of merit for crowded-field observations will not be total scattered power but the broad low-level halo, which redistributes light from one bright star across many faint neighbours, and can damage photometry of resolved stellar populations (SCDD-EE-6) or lensing measurements (SCDD-GG-13) more than broadening the core of the point spread function.
%Stray light in particular is likely to interact with the decision to keep the number of mirrors low, since fewer surfaces means fewer opportunities to place baffles.

% !TEX root = ../main.tex
\section{Detectors and readout}
\label{sec:detectors}

The science goals for \HRI\ (Section~\ref{sec:science}) require exposure times from thousands of seconds on faint galaxies, to millisecond sampling of a bright transiting system, and spanning four orders of magnitude in the number of required pixels. No single readout mode could accommodate all these.
Instead, the detector and its readout electronics offer a family of operating points along a single curve that exchanges illuminated area for cadence.
The same optics, silicon and electronics deliver deep survey imaging, moving target tracking for planetary defence, time-resolved spectrophotometry, or kilohertz-rate photometry, depending only on how the array is clocked.

\HRI\ is thus intended to move between regimes in which angular resolution, field of view, wavelength coverage or time resolution is the limiting resource, and its architecture is arranged so that the limiting resource is selected by the observing mode rather than being fixed permanently by hardware.
This section describes the two focal planes, the physics that limits them, and the family of observing modes they support.

\subsection{The ultraviolet--visible focal plane}

The combination of requirements in Table~\ref{tab:req} points firmly at silicon CMOS.
Three properties of CMOS devices are particularly important.
Windowing allows an arbitrary region of the array to be clocked at high rate while the rest is undisturbed.
Non-destructive readout allows a pixel to be sampled repeatedly during an integration, so that read noise can be beaten down by fitting the resulting ramp rather than by reading slowly.
Programmable analogue gain and ADC resolution allow the same device to serve faint-source and high-dynamic-range observations.
%A CCD could not deliver the desired cadence and low read noise, because it must transfer charge across the device to a small number of output amplifiers.
CMOS devices have less space heritage than CCDs.
However, knowledge of their long-term radiation tolerance is beginning to accumulate from Solar Orbiter and the Chinese Space Agency's Einstein Probe, and the radiation environment at L2 is gradually becoming characterised by Gaia, Euclid and JWST.

\begin{figure}
\centering
\includegraphics[width=0.85\columnwidth]{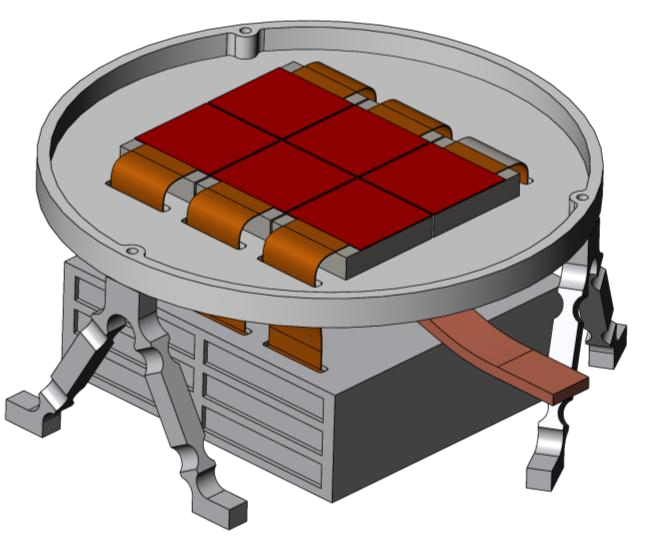}
\caption{The UVIS focal plane assembly.
Six $8192\times8192$ devices tile a $3\times2$ mosaic with small inter-device gaps, giving the $3'\times2'$ field a plate scale of 9.35\,mas per pixel. %; the gaps make the relative detector geometry part of the calibration model rather than a fixed property of the hardware (Section~\ref{sec:mechanical}).
Flexible straps leave the assembly at lower right and connect through heat pipes to a dedicated radiator, thermally isolating the detectors from the optical bench they are mounted on.}
\label{fig:fpa}
\end{figure}

The focal plane is a $3\times2$ mosaic of $8192\times8192$ devices, three-side buttable, at 6.5\,\um\ pitch (Figure~\ref{fig:fpa}).
Each pixel subtends 9.35\,mas. 
Devices of this class are a planned evolution of parts already developed for ESA missions, and pixel pitch is one of the parameters expected to improve on HWO's timescale; the current generation is available at 10\,\um\ in the largest formats and at 7.5\,\um\ in smaller ones.

\subsubsection{Front-end electronics}

Each UVIS front-end electronics unit serves two detectors, so three units read the mosaic.
The design uses a radiation-hardened flash FPGA of a type already flying on ESA instruments.
Each detector presents twelve current mode logic lanes at 1\,Gbit\,s$^{-1}$, and a full $8192\times8192$ frame at 16 bits is 128\,MB and takes 106\,ms to transfer from the sensor to the FPGA.
The FPGA buffers frames into radiation-tolerant DDR memory, which is what makes the fast modes practicable: images can be acquired at high rate and read out to the processing unit more slowly, decoupling acquisition cadence from link bandwidth.

Power is 2.4\,W per sensor during readout, rising to a peak of 6\,W, and falling to about 2\,W while idle or integrating.
The FPGA adds 1.8\,W, and a complete unit dissipates around 7.6\,W.

The choice of serial link matters here.
Over SpaceWire, a full UVIS frame takes 6.4\,s to reach the processing unit on a single link, or 1.6\,s over four.
SpaceFibre offers roughly fifteen times the per-lane rate and is baselined for the science data path, with the front-end memory absorbing the difference between burst acquisition and sustained transfer.
We shall return to this in Section~\ref{sec:electronics}.

\subsection{The near-infrared focal plane}

The NIR channel uses a $3\times2$ mosaic of mercury--cadmium--telluride (MCT) detectors. Each detector will have $4096\times4096$ pixels at a 10\,\um\ pixel pitch, giving $12\,288\times8192$ total pixels. The detectors are required to be sensitive from 900\,nm to 2.2\,\um.

The top-level goal is a read noise of 2.5\,\ep\ with a nominal exposure time of 20\,s. The dark current goal is 0.002\,\ep/pix/s.

The dark current falls as the detector operating temperature is reduced, and increases as the cut-off wavelength of the MCT material is increased. The dark-current goal of 0.002\,\ep/pix/s, with the nominal 2.5\,\um\ MCT cut-off adopted in Section~\ref{sec:science}, could be achieved with an H4RG operating at a temperature of $\sim$90\,K.

\subsubsection{Read noise architecture}
\label{sec:readnoise}
Non-destructive readout allows multiple samples to be taken during an integration and combined as Fowler pairs or by fitting up the ramp. Total noise falls roughly as the square root of the number of samples until charge injected by the act of reading begins to dominate, at which point read noise starts to increase with more samples. Published measurements for an H4RG show the floor being reached at around 64--128 Up-The-Ramp (UTR) reads, with a minimum UTR read noise of $\sim$4.5\,\ep\ when read at 100\,kHz at 90\,K \citep{ives2020}. The number of samples achievable in a fixed exposure is set by the number of outputs and the pixel readout rate.

A standard H4RG-10, sampled at 500\,kHz, could sample approximately 20 UTR reads in a 20\,s exposure with an estimated read noise of $\sim$7.5\,\ep. Doubling the detector outputs to 64 would double the sample count and bring this to $\sim$5\,\ep, however we still wouldn't meet the top-level read noise requirement for the instrument.

An alternative is to apply gain before the read amplifier, rather than average the amplifier's noise away afterwards. Linear-mode avalanche photodiode (APD) arrays multiply the photoelectron signal within the detector, so the effective read noise referred to the input falls as the avalanche bias is raised. Measurements on megapixel-class devices at 75\,K report $<2$\,\ep\ with a single correlated pair with an APD bias of around 10\,V \citep{claveau2024}, consistent with other characterisations of similar arrays \citep{finger2023,owton2025}. HWO would not operate these devices in photon-counting mode; a modest gain is sufficient to reduce, rather than eliminate, the read noise.

An advantage of using APD devices is the reduced power consumption of the system. Reaching the noise requirement with one correlated pair rather than twenty reduces the number of pixel-reads per exposure, and with it the volume of data the front-end electronics must sample and process.

We proposed adopting APD devices for the instrument, as it was the only route we found to meet the top-level read noise requirement.

This choice affects the electronics in two competing ways. Fewer reads per exposure let the readout ASIC run at a lower sample rate, dissipating roughly 100\,mW per detector instead of about 1\,W, a difference of around 5\,W across the mosaic. That saving occurs at the ASIC's 150\,K operating temperature, where radiator area is expensive, so it is worth an order of magnitude more in area than the same wattage would be at room temperature. However, recent results from current APD detectors \citep{owton2025} suggest they will need to operate at a lower temperature to meet the dark current requirement (75\,K rather than 90\,K), and cooling to a lower temperature is thermodynamically more demanding per watt of heat rejected, so this stage carries a disproportionate share of the instrument's cooling burden even though its own dissipation is small. The consequence of this trade-off for radiator sizing is discussed further in Section~\ref{sec:thermal}. The APD architecture is currently expected to require slightly more radiator area than H4RGs, despite dissipating substantially less power overall.

This trade-off is an illustration in the study of how tightly subsystems are bound together, because the cost of this choice appears in the radiator size, three subsystems away from where it is made.

%In the next phase of this study, we intend to investigate early measurements that suggest the dark current in the latest devices is better than expected, which would allow a higher operating temperature.
%Selectively powering down detector outputs might also reduce dissipation when full readout bandwidth is not needed -- at a cost in thermal instrument stability.

\begin{table*}
\centering
\caption{Imager observing modes.
Regions of interest may be contiguous or separated; in fast framerate mode, for example, the allocation can be spent as 36 regions of $1''\times1'$ or as a single contiguous $12''\times3'$ strip.
The long axis of a region is the grism dispersion direction.
For thermal stability, NIR power is intentionally constant, 2.4\,W per detector, in all observing modes 1--4.}
\label{tab:modes}
\begin{tabular}{P{2.6cm}P{3.6cm}P{2.35cm}P{2.35cm}P{4.5cm}}
\toprule
Mode & Sampling & UVIS area & NIR area & Purpose \\
\midrule
1. Full frame &
  up the ramp at 1\,Hz (UVIS), 2\,Hz (NIR); or rolling shutter,
  $10 < t_{\rm exp} < 1000$\,s &
  $3'\times2'$ & $3'\times2'$ &
  Survey and deep imaging, wavefront sensing, most calibration \\
\addlinespace
2. Planetary defence &
  full frame at 1--2\,Hz plus up to six regions at
  $1 < t_{\rm exp} < 1000$\,s &
  ROI $1''\times1'$ or larger & ROI $1''\times1'$ or larger &
  Moving targets tracked against a static field \\
\addlinespace
3. Fast framerate &
  10\,Hz &
  $\le800$ rows ($6''\times1'$) & $\le400$ rows ($6''\times1'$) &
  Time-resolved imaging and grism spectrophotometry \\
\addlinespace
4. Super-fast framerate &
  1\,kHz &
  $\le8$ rows ($60$\,mas$\,\times1'$) & $\le8$ rows ($60$\,mas$\,\times1'$) &
  The brightest and most rapidly varying sources \\
\addlinespace
5. Guiding &
  10\,Hz windowed &
  \multicolumn{2}{P{4.9cm}}{six independent $1''\times1'$ windows per detector} &
  Fine guidance; if guiding in UVIS channel, NIR channel can be used concurrently in any mode \\
\bottomrule
\end{tabular}
\end{table*}

\subsection{Observing modes}
\label{sec:modes}

Table~\ref{tab:modes} sets out the imager observing modes, each corresponds to a class of science identified in Section~\ref{sec:science}.
They are points along a single trade between the number of pixels being read and the rate at which they are read, resulting in intentionally constant heat dissipation from the NIR focal plane in all observing modes 1--4, improving instrument stability.

Mode 1 is the workhorse.
Full-frame readout at 1\,Hz in the UVIS arm and 2\,Hz in the NIR arm, sampled up the ramp, serves the survey programme, supplies the images used for wavefront sensing, and covers most calibration.
The rolling shutter option gives lower noise for exposures between ten and a thousand seconds.

Mode 2 exists for planetary defence (SCDD-SSiC-30) and Solar system science.
It maintains full-frame integration for the static field while reading up to six regions of interest at their own cadence, so that a fast-moving object can be followed without abandoning the astrometric reference frame around it.

Modes 3 and 4 are where the trade becomes explicit.
At 10\,Hz the instrument can read 800 rows per UVIS detector, equivalent to $6''\times1'$.
That allocation can be spent as one contiguous strip along the grism dispersion direction, which is the natural configuration for transit spectrophotometry (SCDD-SSiC-16), or divided into as many as 36 separate $1''\times1'$ regions, which suits monitoring many sources in a crowded field.
Pushing to 1\,kHz reduces the allocation to eight rows, $60\,{\rm mas}\times1'$, enough for a single bright target and the small region around it.

Mode 5 provides fine guidance and runs alongside whatever else the instrument is doing.
Six independent windows per detector are read at 10\,Hz while science integration continues.

One constraint on this flexibility is a design choice rather than a limitation of physics.
Some readout electronics can subdivide a row, allowing regions of interest narrower than the full array width; others cannot.
Baselining sub-row regions would have committed the instrument to a single readout architecture.
We chose instead to define all regions as whole rows, preserving two viable supply routes at the cost of some granularity in how the pixel allocation can be spent.
The trade depends how much of the science programme requires many small boxes rather than a few long strips.
%We will revisit it during the next study phase.

\subsection{Calibration sources}

Photometry at the part-per-million level requires pixel-to-pixel response to be measured, and preferably measured through each science filter so that the flat field corresponds to the passband in use.
Both arms therefore carry calibration lamps that illuminate diffusers mounted in the forward interchange mechanism, with sky photons blocked.
Comparable units have flown on JWST/NIRSpec and on Euclid \citep{hormuth2025}, and the study proposes a NIRSpec-like source for the UVIS arm and a Euclid-like approach for the NIR arm.
The next phase of this study will develop a detailed on-orbit calibration plan.

% !TEX root = ../main.tex
\section{Thermal design}
\label{sec:thermal}

The near-infrared detectors must run at about 75\,K, in a passively cooled instrument, bolted to a telescope at 293\,K.
Everything else in the thermal architecture follows from arranging that.
Thermal dissipation of systems at this range of temperatures is wildly unequal (Table~\ref{tab:thermal}).
The astrometric channel alone accounts for 140\,W, while the entire NIR focal plane dissipates less than a watt.
The coldest systems dissipate least, so their radiator area is set almost entirely by parasitic heat leaking in rather than by the heat they generate.

\begin{figure*}
\centering
\includegraphics[width=0.86\textwidth]{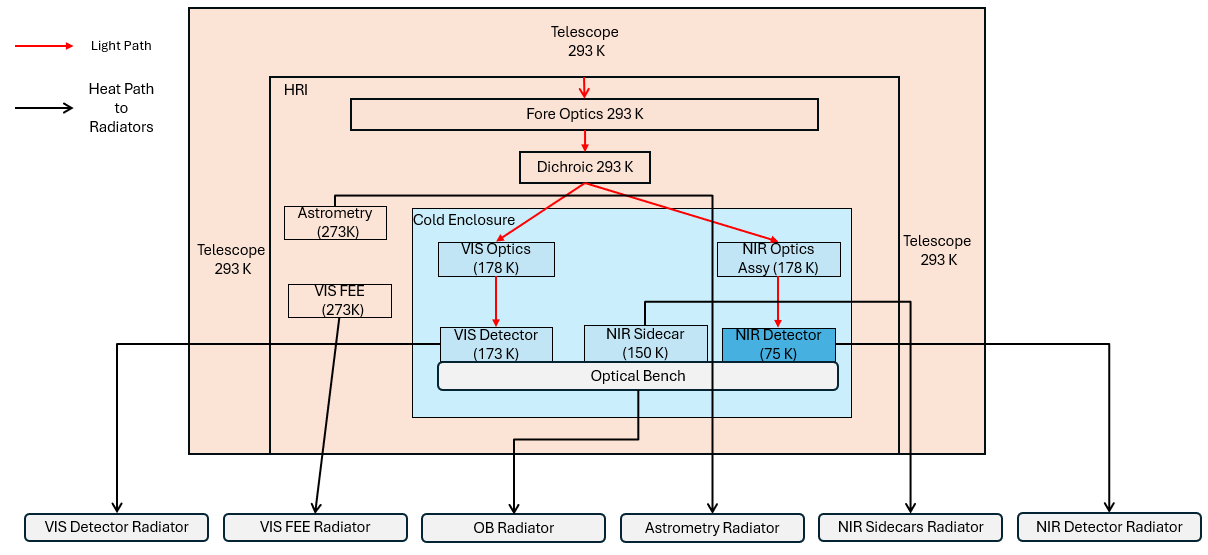}
\caption{Thermal block diagram.
Nesting a cold zone inside a hot zone, rather than giving each subsystem its own enclosure, allows six subsystems spanning 75\,K to 293\,K to be served by one set of passive radiators.
The NIR focal plane and its readout ASICs are isolated from the cold enclosure and cooled separately, because they are the only items that need temperatures below that of the optical bench, and radiator area at those temperatures is the most expensive in the instrument.}
\label{fig:thermalblock}
\end{figure*}

\begin{table}
\centering
\caption{Thermal requirements and dissipations adopted for the study.
Parasitic loads are estimated at 100 per cent of internal dissipation for the cryogenic items and 10 per cent elsewhere.}
\label{tab:thermal}
\begin{tabular}{lccc}
\toprule
Subsystem & Qty & $T$ (K) & $Q_{\rm int}$ (W) \\
\midrule
NIR detectors & 6 & 75 & 0.24 total \\
NIR readout ASICs & 6 & 150 & 0.6 total \\
UVIS detectors & 6 & 220 & 18 total \\
UVIS front-end electronics & 1 & 273 & 18 \\
Astrometric channel & 1 & 293 & 140 \\
Optical bench & 1 & 178 & -- \\
\bottomrule
\end{tabular}
\end{table}

\subsection{Architecture}

Our architecture has three nested temperature zones (Figure~\ref{fig:thermalblock}).
A 293\,K hot zone interfaces with the telescope and carries the foreoptics, forward interchange mechanism, astrometric channel and UVIS front-end electronics.
Inside the hot zone sits a shrouded and insulated 178\,K cold enclosure holding an optical bench with UVIS and NIR arms, plus the UVIS focal plane.
The NIR focal plane and its readout ASICs are thermally isolated and further cooled through separate, dedicated radiators.

Note that the UVIS front-end electronics are mounted in the warm zone, immediately behind the UVIS focal plane to keep the analogue signal path short.
They are the largest dissipator other than the astrometry channel, so it would be thermally expensive to house them in the cold enclosure.
The UVIS optics are kept colder than they need to be.
They dissipate no heat, so there is no thermal cost, but a symmetric, isothermal structure favours alignment stability, as required by the PSF requirements of Section~\ref{sec:science}.
More thermal zones could better isolate the coldest stages and reduce parasitic heating, but would increase mass and make integration more complex.

\subsection{Radiators}

Radiators are mounted on the space-facing side of the hot zone, separated from it by multi-layer insulation.
For a first-order calculation of radiator sizes, we balanced internal and parasitic heat against radiation to an effective sink, taking the radiator emissivity as 0.8 and the sink as a view-factor weighted combination of 70\,per cent deep space at 3\,K and 30\,per cent the cold side of the telescope sunshield at 60\,K.
Conductive paths to each radiator were estimated from the lengths and types of strap and heat pipe involved, ranging from 0.1\,W\,K$^{-1}$ for a copper strap to the NIR focal plane to 20\,W\,K$^{-1}$ for the directly mounted astrometric channel.

\begin{figure}
\centering
\includegraphics[width=0.78\columnwidth]{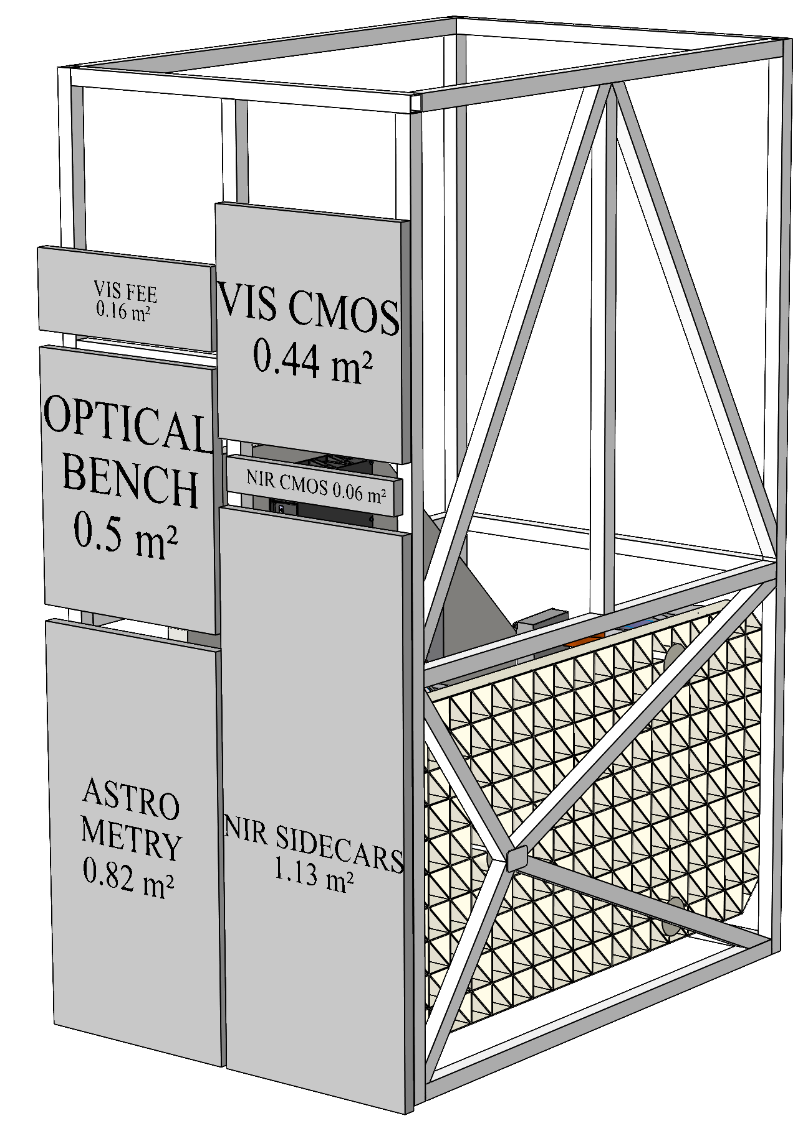}
\caption{Radiator layout on the space-facing face of the \HRI\ enclosure, with the area allocated to each subsystem.
The areas are not proportional to the heat rejected: the NIR readout ASICs dissipate under a watt but take the largest panel at 1.13\,m$^2$ because they radiate from 150\,K, while the astrometric channel dissipates 140\,W into 0.82\,m$^2$ at 293\,K.
This $T^4$ scaling is what makes the detector temperature trade of Section~\ref{sec:readnoise} turn out against intuition.}
\label{fig:radiators}
\end{figure}

Radiator sizing was run for three cases, applying multipliers of $\pm20$ per cent to internal dissipation and $\pm50$ per cent to the linear conductors.
All three fit within the 4.28\,m$^2$ requirement.
In the nominal layout (Figure~\ref{fig:radiators}) the astrometric channel takes 0.82\,m$^2$, the UVIS detectors 0.44\,m$^2$, the optical bench 0.50\,m$^2$, the UVIS front-end electronics 0.16\,m$^2$, and the NIR readout ASICs 1.13\,m$^2$.
Headroom should not yet be regarded as flight thermal margin.
Spacecraft attitude history, sunshield temperatures, interface conductances and real component dissipations must all be included before the design can be closed.

We then built a coarse instrument-level model in the ESATAN Thermal Modelling Suite to check that the parasitic assumptions are credible.
Predicted component temperatures agree with the sizing assumptions to within 8\,K for all subsystems, and radiator temperatures to within 7\,K, with the model running slightly colder than assumed in every case except the optical bench.
This level of agreement is acceptable for architecture-level trades at this maturity without indicating that either calculation is more accurate than the other.

\subsection{Heaters}

Temperature sensing uses platinum resistance thermometers distributed through the instrument, and heaters follow qualified designs, both similar to those already flown on Euclid.
Decontamination heaters are foreseen, driven by ground command or on a schedule, with current monitoring on each circuit so that open and short circuits can be detected.
Survival heater control has not been allocated between instrument and spacecraft.

% !TEX root = ../main.tex
\section{Mechanical design and accommodation}
\label{sec:mechanical}

The UVIS and NIR arms share a single cold optical bench carrying their mirrors, filter wheels, focus stages and focal planes.
Separate benches would let each arm be optimised independently but would double the alignment problem and make co-registration between the channels a system-level rather than a bench-level activity.
Since the purpose of the dichroic in the forward interchange mechanism is that the two arms observe the same field at the same time (Section~\ref{sec:simultaneity}), their relative alignment is a science requirement rather than a convenience, and a common bench is the natural way to hold it.

\begin{figure}
\centering
\includegraphics[width=\columnwidth]{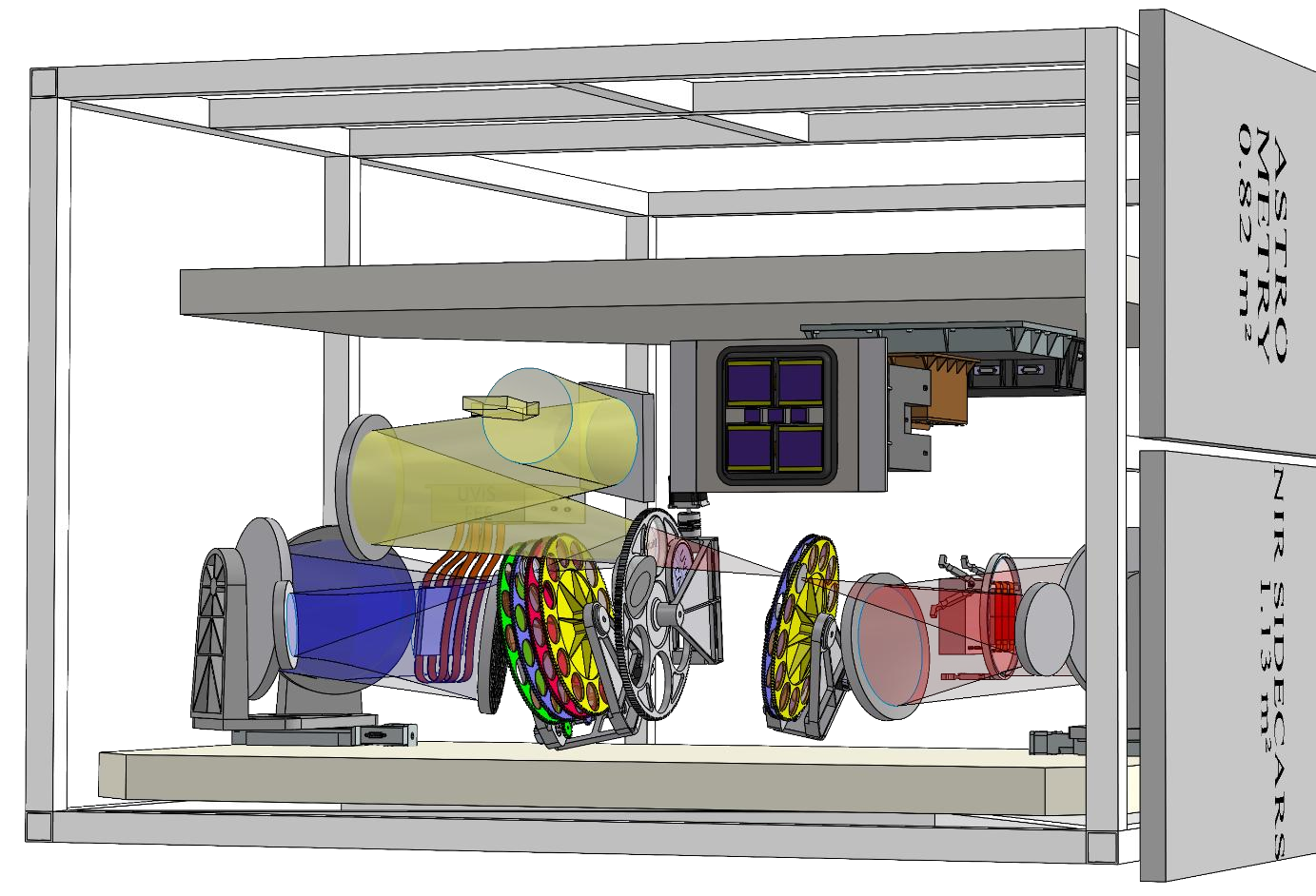}
\caption{Accommodation within the EAC-1 volume requirement, with the astrometric channel positioned above the UVIS and NIR arms and beam envelopes shown in colour.
Stacking the channels vertically allows an instrument with three imaging trains to fit inside the required envelope, and it places the astrometric detector head, data processing unit and power supply unit on the room temperature bench, outside the cold enclosure that surrounds the imager optics.}
\label{fig:packaging}
\end{figure}

Warm and dissipating items sit on a second, room temperature bench: the forward interchange mechanism and its mirrors, the UVIS front-end electronics and the astrometric channel.
Figure~\ref{fig:packaging} shows the arrangement, with the astrometric channel stacked above the two imager arms inside the shared volume.
Its optical axis can be brought as close as 300\,mm to the axis entering the forward interchange mechanism, though the relative placement of the two was not fixed during the study and should be traded in the next phase to optimise the enclosure.

%Mount stiffness, compliance and thermal conductance have to be traded together against the optical tolerance budget, and that trade cannot be done properly until the budget exists in the wavefront terms described in Section~\ref{sec:optics-performance}.

%Focal plane geometry becomes part of the calibration model rather than a fixed property of the hardware.
%A six-device mosaic must sit accurately in the image plane while accommodating differential contraction between the detectors, their packages and the bench beneath them, and the small gaps between devices, useful for field coverage, mean the relative detector positions have to be solved for.
%Metrology of the assembled focal plane and a stable in-flight distortion solution will therefore be significant commissioning activities.

%Servicing shapes the enclosure.
HWO is intended to be robotically serviceable at L2 \citep{arenberg25,liu2026}, and servicing accommodation is already shaping the EAC designs. 
The face of the volume that is externally accessible and views cold space is the face carrying the radiators, so thermal connections are made with heat pipes and flexible straps to allow the module to be removed.

The 285\,kg imager mass comfortably meets the 450\,kg requirement (Table~\ref{tab:mass}).
The mass is 415\,kg including the astrometric channel; NASA's baseline does not cover that channel.
Harness, radiator support frames, bipods and the enclosure-to-spacecraft interface are not yet included.
There is nonetheless plenty of margin for additional optical elements if needed (see Section~\ref{sec:optics-performance}).

\begin{table}
\centering
\caption{Mass budget summary.
The 450\,kg requirement excludes the astrometric channel.
Harness, radiator support structure and bipods are not yet included.}
\label{tab:mass}
\begin{tabular}{lr}
\toprule
Subsystem & Mass (kg) \\
\midrule
Structure (benches, enclosures, frame) & 164.8 \\
Thermal (radiators, heaters, sensors) & 28.3 \\
UVIS channel & 27.5 \\
NIR channel & 22.6 \\
Control and data processing electronics & 22.5 \\
Mechanism control & 10.0 \\
Power supply unit & 6.0 \\
Fine guidance sensor & 3.0 \\
\midrule
\textbf{Imager total} & \textbf{284.7} \\
Astrometric channel & 130.0 \\
\textbf{Total including astrometry} & \textbf{414.7} \\
\midrule
Requirement (excluding astrometry) & 450.0 \\
Margin & 37 per cent \\
\bottomrule
\end{tabular}
\end{table}

% !TEX root = ../main.tex
\section{Electronics, data handling and resources}
\label{sec:electronics}

\subsection{A single processing unit}

The instrument electronics are organised around one decision: to merge the instrument control unit and the data processing unit into a single Control and Data Processing Unit (CDPU) that also handles mechanism control, heater control and health monitoring.
The CDPU is the sole interface to the spacecraft for commanding, telemetry and science data.

The argument for consolidation is interface count.
Separate control and processing units require an interface between them, each requires its own spacecraft interface and its own interface control document, and each adds harness.
Folding mechanism and heater control into the same unit removes several more.
Against this, a merged unit is a larger and less independently testable item, and it concentrates function in a way that makes the redundancy scheme more important.
The precedent is Euclid \citep{cropper2025}, where merged units of comparable scope flew successfully, and PLATO \citep{rauer2024}.

The astrometric channel retains its own data processing unit and memory, reflecting its independent science objectives and its internal data volumes, but is commanded through the CDPU so that the instrument still presents a single face to the spacecraft.

\begin{figure}
\centering
\includegraphics[width=\columnwidth]{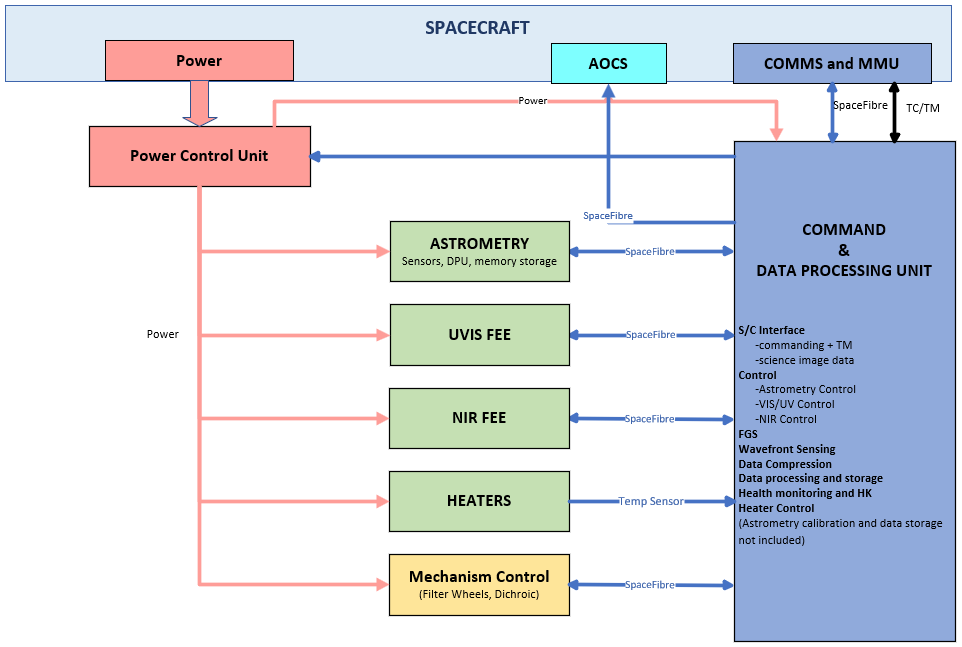}
\caption{Instrument electronics architecture.
Consolidating command, data processing, mechanism control and heater control into a single CDPU leaves the instrument with one spacecraft interface instead of several, which simplifies the harness, serviceability and documentation, at the cost of a larger and less independently-testable unit.
Science data reach the CDPU over SpaceFibre from the UVIS and NIR front-end electronics and from the astrometric channel, whose own data processor is retained because of its independent science objectives and internal data volumes.}
\label{fig:electronics}
\end{figure}

The CDPU comprises six boards on an internal backplane: a processor board running the flight software and fault management, a data processing board implementing the SpaceFibre links and lossless compression, an input/output board carrying the external interfaces and the thermometry front end, a mass memory board with single-error-correcting and double-error-detecting protection, a power board, and the backplane itself carrying power, clock and inter-board communication.
Redundancy is foreseen for the CDPU, the power supply unit, the mechanism controller, the fine guidance function, the astrometric data processor and the mechanism motor windings, in line with NASA practice for payloads of this class.

\subsection{Power}

The 162\,W mean power draw meets the 183\,W requirement (Table~\ref{tab:power}) but, with only 12\,per cent margin, this will require active management.
This does not include the astrometric channel (166\,W), for which no NASA requirement currently exists.

We have assumed a power supply unit with 75 per cent conversion efficiency to provide the imager's 121\,W.
Margin may come from modern space-qualified converters, which routinely achieve 85 to 90 per cent efficiency.
A proper duty-cycle analysis, which does not yet exist, may reveal further margin in the time-averaged figure since mechanisms actuate rarely and heaters operate intermittently.
Our design philosophy also includes permanently operating UVIS front-end electronics, to maintain thermal stability, but could be relaxed.
Consolidating more detectors per front-end FPGA would also reduce unit count and total dissipation.

\begin{table}
\centering
\caption{Power budget summary, average values.
Requirements exclude the astrometric channel.}
\label{tab:power}
\begin{tabular}{lr}
\toprule
Subsystem & Power (W) \\
\midrule
Control and data processing electronics & 60.0 \\
UVIS channel & 35.0 \\
Mechanism control & 10.0 \\
NIR channel & 8.1 \\
Fine guidance sensor & 8.0 \\
\midrule
\textbf{Imager total} & \textbf{121.1} \\
PSU input at 75 per cent efficiency & 161.5 \\
Requirement & 183.3 \\
\midrule
Astrometric channel & 207.0 \\
Total including astrometry & 328.1 \\
\bottomrule
\end{tabular}
\end{table}

\subsection{SpaceFibre and data rates}

Science data move over SpaceFibre \citep{parkes2015} rather than SpaceWire.
The bandwidth argument is straightforward: at SpaceWire rates a single full UVIS frame would take seconds to transfer (Section~\ref{sec:detectors}), and bound the achievable duty cycle in fast observing modes even with front-end buffering.
SpaceFibre also offers virtual channels, so that time-critical guidance data and housekeeping can share a physical link with bulk science data while retaining guaranteed latency.
Heritage for SpaceFibre is accumulating and should be well established by the time \HRI\ is built.
The margin it provides for growth in data rate is worth the residual risk.

A full UVIS frame is 5.6\,Gbit at 14 bits per pixel across six devices, and a full NIR frame 1.6\,Gbit at 16 bits.
Applying the average compression ratios achieved by Euclid (2.6 for VIS and 2.4 for NIR; VIS team, priv.\ comm.), these become 2.16\,Gbit and 0.67\,Gbit respectively.
Mass memory sizing has not been fixed, since it depends on downlink cadence and duration, which are mission level parameters not yet available.

\subsection{Electromagnetic compatibility}

A high-rate imaging instrument has to control the electromagnetic consequences of its own readout.
Detector clocks, multi-gigabit serial links and fast analogue-to-digital conversion all generate switching transients, and several large detectors run at once.
The partition adopted here helps: the front-end electronics perform only what must happen close to the detector, which is bias generation, analogue buffering, timing and conversion, while compression, packetisation, memory management and instrument-level decisions stay in the CDPU.
That keeps the long, fast connections digital rather than analogue, and it keeps the sensitive analogue paths short and locally screened.
A full compatibility analysis has not been done and will be needed before the electronics design is frozen.

% !TEX root = ../main.tex
\section{Discussion}
\label{sec:discussion}

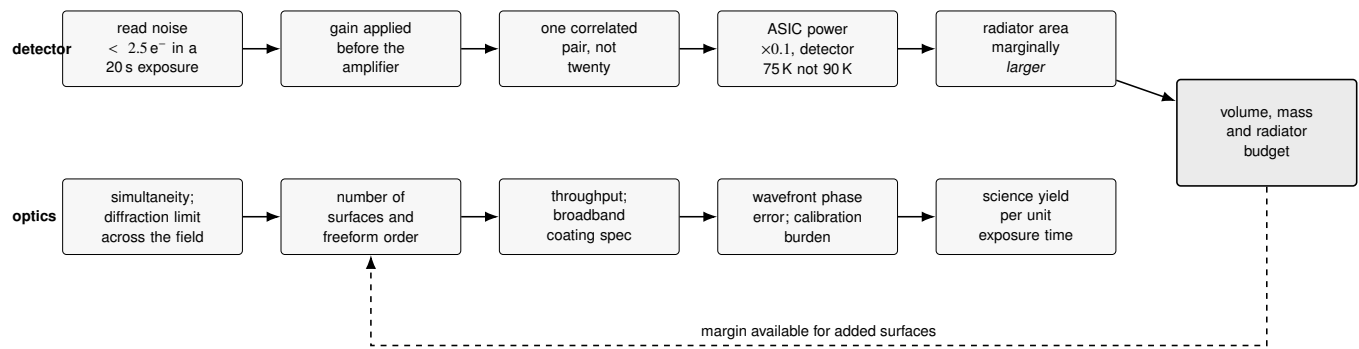
\begin{figure*}
\centering
\resizebox{\textwidth}{!}{%
\begin{tikzpicture}[
  font=\sffamily\small,
  box/.style={draw, rounded corners=2pt, align=center, minimum height=1.35cm,
              text width=3.0cm, inner sep=3pt, fill=black!3},
  res/.style={draw, thick, rounded corners=2pt, align=center,
              minimum height=1.9cm, text width=3.0cm, inner sep=3pt,
              fill=black!8},
  ar/.style={-{Latex[length=2.5mm]}, thick},
  dashar/.style={-{Latex[length=2.5mm]}, thick, dashed}
]

% ---- detector chain -------------------------------------------------
\node[box] (a1) at (0,3.0)  {read noise\\ $<2.5$\,\ep\ in a\\ 20\,s exposure};
\node[box] (a2) at (3.9,3.0) {gain applied\\ before the\\ amplifier};
\node[box] (a3) at (7.8,3.0) {one correlated\\ pair, not\\ twenty};
\node[box] (a4) at (11.7,3.0) {ASIC power\\ $\times 0.1$, detector\\ 75\,K not 90\,K};
\node[box] (a5) at (15.6,3.0) {radiator area\\ marginally\\ \emph{larger}};

\draw[ar] (a1) -- (a2);
\draw[ar] (a2) -- (a3);
\draw[ar] (a3) -- (a4);
\draw[ar] (a4) -- (a5);

% ---- optical chain --------------------------------------------------
\node[box] (b1) at (0,0)  {simultaneity;\\ diffraction limit\\ across the field};
\node[box] (b2) at (3.9,0) {number of\\ surfaces and\\ freeform order};
\node[box] (b3) at (7.8,0) {throughput;\\ broadband\\ coating spec};
\node[box] (b4) at (11.7,0) {wavefront phase\\ error; calibration\\ burden};
\node[box] (b5) at (15.6,0) {science yield\\ per unit\\ exposure time};

\draw[ar] (b1) -- (b2);
\draw[ar] (b2) -- (b3);
\draw[ar] (b3) -- (b4);
\draw[ar] (b4) -- (b5);

% ---- shared resource ------------------------------------------------
\node[res] (r) at (19.9,1.5) {volume, mass\\ and radiator\\ budget};
\draw[ar] (a5) -- (r);
\draw[dashar] (r.south) -- (19.9,-2.3) --
      node[above,font=\sffamily\footnotesize]
      {margin available for added surfaces} (3.9,-2.3) -- (b2.south);

% ---- labels ---------------------------------------------------------
\node[anchor=west,font=\sffamily\bfseries\small] at (-2.6,3.0) {detector};
\node[anchor=west,font=\sffamily\bfseries\small] at (-2.6,0)   {optics};

\end{tikzpicture}}
\caption{The two dependency chains that dominated the study.
A requirement stated at the left of each chain is paid for at the right, and the two chains compete for the same volume, mass and radiator budget.
Neither chain can be closed by the subsystem in which its requirement originates.}
\label{fig:coupling}
\end{figure*}

Our concurrent design approach revealed dependencies between subsystems at an early stage, by having optical, thermal, mechanical, detector and electronics engineers in one room and NASA HTMPO on a daily call.
Since the instrument design will need to change as the EACs evolve, the most useful result of this study may be those dependencies.
%The science requirements in Section~\ref{sec:science} are not independent, and neither are the subsystems that meet them.
Binding constraints are rarely internal to one subsystem or one section of this paper; they arise where a choice made in one place is paid for in another.
The two chains that dominated this study are summarised in Figure~\ref{fig:coupling}.

\subsection{The detector requirements chain}

A read noise requirement is a statement about photon counting.
It becomes, in sequence, a choice of readout architecture; a number of pixel samples per exposure; a data rate through the front-end electronics; a dissipation in the readout ASIC; a detector operating temperature; a radiator area at that temperature; and finally a constraint on how the instrument is divided into thermal zones and therefore on the mechanical layout.

Section~\ref{sec:readnoise} followed one example of this.
The architecture that meets the read noise requirement reduces ASIC dissipation by roughly a factor of ten, but still ends up needing slightly more radiator area, because it also lowers the detector operating temperature from 90\,K to 75\,K and radiated power scales as $T^4$.
Nothing about that outcome is visible from the detector specification.
It emerges only when the chain is followed through, and it is the kind of result that a subsystem-by-subsystem study might systematically miss.
This trade is still flexible, as the point design retains considerable margin in mass and volume that could be used to add radiator area or an additional thermal zone.

\subsection{The optical requirements chain}

The optical chain runs the other way, from a science requirement to a science outcome, through a technical decision in the optical design.

Simultaneity across the band requires a dichroic.
A dichroic costs a surface but permits two arms built by different routes, so the total surface count stays low while each arm is properly matched to its band (Section~\ref{sec:optics}).
Low surface count buys throughput, a shorter list of broadband coatings to qualify, fewer mounts to hold stable, and fewer phase terms in the wavefront error budget (Section~\ref{sec:coatings}).
All of that converts into science yield per unit exposure time, which is the currency a facility observatory actually trades in.

This chain also contains the study's principal open item.
Making the design diffraction limited across the whole field, rather than at the points where it was optimised, may require a slower system with more separated surfaces, or one or two additional surfaces.
The can be easily accommodated through existing margin in mass and volume. 
But each surface runs the chain backwards: less throughput, another coating, another mount, another phase term.
Whether the exchange is worth making depends how much the wavefront-sensing budget is degraded by a dielectric coating phase at $12^\circ$ incidence, and how much point spread function non-uniformity the gravitational lensing and transit programmes can tolerate. 
These have not yet been measured, but are well-posed questions with a bounded answer, that can be investigated next.

% !TEX root = ../main.tex
\section{Conclusions}
\label{sec:conclusions}

The study set out to establish whether a wide-band, near diffraction-limited imager meeting HWO's expectations is feasible, and concluded that it is.
The design fits within requirements for mass, volume and power; meets requirements for the field of view, wavelength coverage and sampling; and provides the fine guidance, wavefront sensing and pupil imaging functions the observatory needs from it.
At locations in the field where the optical design was optimised, it is diffraction-limited at 850\,nm in the NIR arm, and close to diffraction-limited at 420\,nm in the UVIS arm, exceeding requirements set by NASA.
Extending that performance across the full field is unfinished, and may require one or two additional surfaces or a slower system with more separation between mirrors.

Four choices define the architecture, each coupled to science goals rather than being an engineering convenience.

\begin{my_itemize}
\item Splitting the beam into separate ultraviolet--visible and near-infrared trains delivers the simultaneous multiband coverage required by Venus monitoring, transit spectrophotometry and the Cepheid distance ladder.
This uses a dichroic, which reduces throughput, but the dichroic can be removed and allows downstream optics to use more reflective coatings optimised for a narrower band, and more appropriate materials: polished ultra-low expansion glass where the diffraction limit is 420\,nm, diamond-turned aluminium where it is 850\,nm.

\item Collimating with only two mirrors before the dichroic instead of four, and reaching the focal planes with six or seven reflections in total, maintains throughout across a full 200\,nm to 2.2\,\um\ wavelength range.
Every surface upstream of the dichroic carries a coating that must work across more than a decade in wavelength, and every surface costs throughput, adds a qualification item, contributes to the stability budget and puts another phase term into the wavefront error budget on which the whole observatory depends.

\item Readout electronics selectable on orbit can trade illuminated area for cadence across roughly three orders of magnitude. 
This allows one set of hardware to serve deep survey imaging, moving-target tracking for planetary defence, time-resolved spectrophotometry of transiting planets, and kilohertz-rate photometry.
The observer, rather than the hardware, can chose which of angular resolution, field of view or time resolution is the limiting resource for any exposure.

\item The staged thermal architecture holds six subsystems at temperatures from 75\,K to 293\,K within a 4.28\,m$^2$ radiator requirement, by putting the dissipating electronics outside the cold enclosure and isolating the cryogenic focal plane inside it.\\
\end{my_itemize}

NASA requires all critical technologies for HWO to reach technology readiness level 5, meaning breadboarded components validated in a relevant environment, by Mission Concept Review at the end of 2029 \citep{bolcar26}.
The study assessed every item in the \HRI\ product tree against that target using a questionnaire based on ECSS practice.
Most items score well, because most of them have close relatives in flight.
Where scores are low they are generally low for a specific and addressable reason rather than because the underlying technology is immature: a component change from a qualified design, or an existing capability that has not been demonstrated in the exact configuration required.

The UK Space Agency has funded a continuation of this study.
The next phase will retain the concurrent design approach, to combine the best features of this and a previously separate UK feasibility study \citep{skottfelt26} into a single instrument concept, while adding international partners.
% ---------------------------------------------------------------------

\section*{Acknowledgements}

The study described here was funded by the UK Space Agency, through grants UKRI3163 and UKRI5725.
The astrometric channel is developed by a separately-funded French team.
We thank colleagues at NASA's HWO Technology Maturation Project Office for their engagement throughout, including every day of our Concurrent Design studies, and the staff of the Concurrent Design Facility at the University of Portsmouth.

\section*{Data Availability}

No new observational data were generated by this study.
The instrument design parameters quoted in this paper are given in full in the tables and figures.

\bibliographystyle{rasti}
\bibliography{refs}

@book{decadal2021,
  author    = {{National Academies of Sciences, Engineering, and Medicine}},
  title     = {{Pathways to Discovery in Astronomy and Astrophysics for the 2020s}},
  publisher = {The National Academies Press},
  address   = {Washington, DC},
  year      = {2021},
  doi       = {10.17226/26141}
}

@article{luvoir2019,
  author  = {{The LUVOIR Team}},
  title   = {{The LUVOIR Mission Concept Study Final Report}},
  journal = {ArXiv e-prints},
  year    = {2019},
  eprint  = {1912.06219},
  archivePrefix = {arXiv}
}

@inproceedings{scowen2025,
  author    = {Scowen, P.~A. and Bolcar, M.~R. and Zhao, F.},
  title     = {{Technology roadmapping and maturation planning to enable UV-Vis astrophysics with the Habitable Worlds Observatory}},
  booktitle = {UV/Optical/IR Space Telescopes and Instruments: Innovative Technologies and Concepts XII},
  series    = {Proc. SPIE},
  volume    = {13623},
  pages     = {1362302},
  year      = {2025},
  doi       = {10.1117/12.3063289}
}

@article{vaneylen2025,
  author  = {Van Eylen, V. and Massey, R. and others},
  title   = {{Transformational astrophysics and exoplanet science with Habitable Worlds Observatory's High Resolution Imager}},
  journal = {ArXiv e-prints},
  year    = {2025},
  eprint  = {2512.16416},
  archivePrefix = {arXiv}
}

@article{amiaux2025,
  author  = {Amiaux, J. and others},
  title   = {{System analysis for a high-precision high-accuracy astrometric instrument for HWO}},
  journal = {ArXiv e-prints},
  year    = {2025},
  eprint  = {2511.07113},
  archivePrefix = {arXiv}
}

@inproceedings{malbet2026,
  author    = {Malbet, F. and others},
  title     = {{Very high precision astrometry for exoplanets and dark matter with the Habitable Worlds Observatory}},
  booktitle = {Astronomical Society of the Pacific Conference Series},
  volume    = {543},
  pages     = {65},
  year      = {2026},
  eprint    = {2510.18920},
  archivePrefix = {arXiv}
}

@article{malbet2021,
  author  = {Malbet, F. and others},
  title   = {{Faint objects in motion: the new frontier of high precision astrometry}},
  journal = {Experimental Astronomy},
  volume  = {51},
  pages   = {845},
  year    = {2021}
}

@article{boehm2017,
  author  = {Boehm, C. and others},
  title   = {{Theia: Faint objects in motion or the new astrometry frontier}},
  journal = {ArXiv e-prints},
  year    = {2017},
  eprint  = {1707.01348},
  archivePrefix = {arXiv}
}

@article{batalha2019,
  author  = {Batalha, N.~E. and Lewis, T. and Fortney, J.~J. and Batalha, N.~M. and
             Kempton, E. and Lewis, N.~K. and Line, M.~R.},
  title   = {{The precision of mass measurements required for robust atmospheric characterization of transiting exoplanets}},
  journal = {ApJ},
  volume  = {885},
  pages   = {L25},
  year    = {2019}
}

@article{wakeford2025,
  author  = {Wakeford, H.~R. and others},
  title   = {{Characterizing the dynamics and chemistry of transiting exoplanets with the Habitable World Observatory}},
  journal = {ArXiv e-prints},
  year    = {2025},
  eprint  = {2506.22839},
  archivePrefix = {arXiv}
}

@article{cartwright2025,
  author  = {Cartwright, R.~J. and Quick, L. and Neveu, M. and Becker, T.~M. and
             Raut, U. and Castillo-Rogez, J. and Craft, K.~L. and Villanueva, G.~L.},
  title   = {{Assessing ocean world habitability with HWO}},
  journal = {ArXiv e-prints},
  year    = {2025},
  eprint  = {2507.07128},
  archivePrefix = {arXiv}
}

@article{adhikari2025,
  author  = {Adhikari, S. and Banerjee, A. and Boddy, K.~K. and others},
  title   = {{Astrophysical tests of dark matter self-interactions}},
  journal = {Reviews of Modern Physics},
  volume  = {97},
  pages   = {045004},
  year    = {2025}
}

@article{battaglia2022,
  author  = {Battaglia, G. and Nipoti, C.},
  title   = {{Stellar dynamics and dark matter in Local Group dwarf galaxies}},
  journal = {Nature Astronomy},
  volume  = {6},
  pages   = {659},
  year    = {2022}
}

@article{vegetti2024,
  author  = {Vegetti, S. and Birrer, S. and Despali, G. and others},
  title   = {{Strong gravitational lensing as a probe of dark matter}},
  journal = {Space Science Reviews},
  volume  = {220},
  pages   = {58},
  year    = {2024},
  eprint  = {2306.11781},
  archivePrefix = {arXiv}
}

@article{cropper2025,
  author  = {Cropper, M.~S. and others},
  title   = {{Euclid: the VIS instrument}},
  journal = {A\&A},
  volume  = {697},
  pages   = {A2},
  year    = {2025}
}

@inproceedings{ives2020,
  author    = {Ives, D.~J. and Alvarez, D. and Bezawada, N. and George, E. and Serra, B.},
  title     = {{Characterisation, performance and operational aspects of the H4RG-15 near infrared detectors for the MOONS instrument}},
  booktitle = {X-Ray, Optical, and Infrared Detectors for Astronomy IX},
  series    = {Proc. SPIE},
  volume    = {11454},
  pages     = {114540L},
  year      = {2020},
  doi       = {10.1117/12.2562408}
}

@article{claveau2024,
  author  = {Claveau, C.-A. and Bottom, M. and Jacobson, S. and Hodapp, K. and Huber, G. and
             Newland, M. and Walk, A. and Loose, M. and Baker, I. and Zemaityte, E. and
             Hicks, M. and Barnes, K. and Powell, R. and Bradley, R. and Moore, E.},
  title   = {{Progress towards a megapixel linear-mode avalanche photodiode array for ultra-low background shortwave infrared astronomy}},
  journal = {ArXiv e-prints},
  year    = {2024},
  eprint  = {2411.09185},
  archivePrefix = {arXiv}
}

@article{finger2023,
  author  = {Finger, G. and others},
  title   = {{Scientific detector workshop 2022 on-sky performance verification of near-infrared e-APD technology for wavefront sensing and demonstration of e-APD pixel performance to improve the sensitivity of large science focal planes}},
  journal = {Astronomische Nachrichten},
  volume  = {344},
  number  = {8--9},
  year    = {2023},
  doi     = {10.1002/asna.20230069}
}

@inproceedings{owton2025,
  author    = {Owton, D. and Baker, I.~M. and Bradford, A. and Isgar, V. and Hicks, M. and Maxey, C.~D.},
  title     = {{Leonardo large area avalanche photo-diode arrays}},
  booktitle = {Infrared Sensors, Devices, and Applications XV},
  series    = {Proc. SPIE},
  year      = {2025},
  doi       = {10.1117/12.3065654}
}

@inproceedings{zane2016,
  author    = {Zane, S. and Winter, B. and Theobalds, C. and Theodorou, T. and
               Pinchera, M. and Muleri, F. and Spada, F. and Sgro, C. and
               Zanetti, D. and Feng, H. and Del Monte, E. and Rubini, A. and
               Soffitta, P. and Bellazzini, R. and Bozzo, E. and Burwitz, V. and
               Castro Tirado, A.~J. and Costa, E. and Gburek, S. and
               Goosmann, R. and Karas, V. and Matt, G. and Nandra, K. and
               Pearce, M. and Poutanen, J. and Reglero, V. and Sabau, M.~D. and
               Santangelo, A. and Tagliaferri, G. and Tenzer, C. and
               Vink, J. and Weisskopf, M.},
  title     = {{The on-board calibration system of the X-ray Imaging Polarimetry Explorer (XIPE)}},
  booktitle = {Space Telescopes and Instrumentation 2016: Ultraviolet to Gamma Ray},
  series    = {Proc. SPIE},
  volume    = {9905},
  pages     = {99054H},
  year      = {2016},
  doi       = {10.1117/12.2233448}
}

@inproceedings{stoneking2025,
  author    = {Stoneking, E. and Hsu, O.~C. and Welter, G. and Lewis, L. and
               Carter, B. and Doerr, B. and Dongmo, J.-E. and Heron, M. and
               Shapiro, J.},
  title     = {{The Nancy Grace Roman Space Telescope's Attitude Control System}},
  booktitle = {47th Annual AAS Guidance, Navigation and Control Conference},
  address   = {Broomfield, CO},
  year      = {2025},
  url       = {https://ntrs.nasa.gov/citations/20240016454}
}

@article{hormuth2025,
  author  = {{Euclid Collaboration} and Hormuth, F. and Jahnke, K. and others},
  title   = {{Euclid: IV. The NISP Calibration Unit}},
  journal = {A\&A},
  volume  = {697},
  pages   = {A4},
  year    = {2025}
}

@article{zodipy2024,
  author  = {San, M.},
  title   = {{ZodiPy: A Python package for zodiacal light simulations}},
  journal = {Journal of Open Source Software},
  volume  = {9},
  number  = {96},
  pages   = {6648},
  year    = {2024},
  doi     = {10.21105/joss.06648}
}

@article{obrien2026,
  author  = {O'Brien, R. and Arendt, R.~G. and Windhorst, R.~A. and others},
  title   = {{SKYSURF-11: A new zodiacal light model optimized for optical wavelengths}},
  journal = {ApJ},
  volume  = {1000},
  pages   = {6},
  year    = {2026}
}

@misc{postman_etc,
  author = {Postman, M.},
  title  = {{Computing signal-to-noise ratio for astronomical point sources}},
  note   = {\url{https://jt-astro.science/luvoir_simtools/hdi_etc/SNR_equation.pdf}},
  year   = {2019}
}

@article{scicases2026,
  author  = {Dressing, C.~D. and others},
  title   = {{Early Exploration of the Scientific Discovery Space for the Habitable Worlds Observatory}},
  journal = {ArXiv e-prints},
  year    = {2026},
  eprint  = {2608.11294},
  archivePrefix = {arXiv}
}

@article{gaudi2020,
  author  = {Gaudi, B.~S. and others},
  title   = {{The Habitable Exoplanet Observatory (HabEx) Mission Concept Study Final Report}},
  journal = {ArXiv e-prints},
  year    = {2020},
  eprint  = {2001.06683},
  archivePrefix = {arXiv}
}

@article{riess2022,
  author  = {Riess, A.~G. and others},
  title   = {{A comprehensive measurement of the local value of the Hubble constant with 1\,km\,s$^{-1}$\,Mpc$^{-1}$ uncertainty from the Hubble Space Telescope and the SH0ES team}},
  journal = {ApJ},
  volume  = {934},
  pages   = {L7},
  year    = {2022}
}

@article{freedman2019,
  author  = {Freedman, W.~L. and others},
  title   = {{The Carnegie-Chicago Hubble Program. VIII. An independent determination of the Hubble constant based on the tip of the red giant branch}},
  journal = {ApJ},
  volume  = {882},
  pages   = {34},
  year    = {2019}
}

@article{mcelwain2023,
  author  = {McElwain, M.~W. and others},
  title   = {{The James Webb Space Telescope mission: optical telescope element design, development, and performance}},
  journal = {PASP},
  volume  = {135},
  pages   = {058001},
  year    = {2023}
}

@inproceedings{acton2012,
  author    = {Acton, D.~S. and others},
  title     = {{Wavefront sensing and controls for the James Webb Space Telescope}},
  booktitle = {Space Telescopes and Instrumentation 2012: Optical, Infrared, and Millimeter Wave},
  series    = {Proc. SPIE},
  volume    = {8442},
  pages     = {84422H},
  year      = {2012}
}

@article{wright2023,
  author  = {Wright, G.~S. and others},
  title   = {{The Mid-Infrared Instrument for JWST and its in-flight performance}},
  journal = {PASP},
  volume  = {135},
  pages   = {048003},
  year    = {2023}
}

@inproceedings{parkes2015,
  author    = {Parkes, S. and others},
  title     = {{SpaceFibre: A multi-Gigabit/s interconnect for spacecraft onboard data handling}},
  booktitle = {2015 IEEE Aerospace Conference},
  pages     = {1--13},
  year      = {2015},
  doi       = {10.1109/AERO.2015.7119317}
}

@article{windhorst2022,
  author  = {Windhorst, R.~A. and others},
  title   = {{SKYSURF: Constraints on zodiacal light and extragalactic background light through panchromatic HST all-sky surface-brightness measurements. I. Survey overview and methods}},
  journal = {AJ},
  volume  = {164},
  pages   = {141},
  year    = {2022}
}

@article{rieke2023,
  author  = {Rieke, M.~J. and Kelly, D.~M. and Misselt, K. and Stansberry, J. and
             Boyer, M. and Beatty, T. and Egami, E. and Florian, M. and Greene, T.~P. and
             Hainline, K. and others},
  title   = {{Performance of NIRCam on JWST in Flight}},
  journal = {PASP},
  volume  = {135},
  pages   = {028001},
  year    = {2023},
  doi     = {10.1088/1538-3873/acac53},
  eprint  = {2212.12069},
  archivePrefix = {arXiv}
}

@inproceedings{vila2025,
  author    = {Vila, M.~B. and Barringer, B. and Bond, N. and Chayer, P. and
               DiFelice, A. and Dupuis, J. and Holfeltz, S. and Lambros, S.~D. and
               Marrione, A. and Moller, C.~D. and Nelan, E. and Rowlands, N. and
               Sohn, S.~T. and Warner, G. and Yeates, J. and Zhou, J.},
  title     = {{JWST Fine Guidance Sensor Commissioning and Performance During Operations}},
  booktitle = {18th International Conference on Space Operations},
  address   = {Montreal, Canada},
  year      = {2025},
  url       = {https://ntrs.nasa.gov/citations/20250002384},
  note      = {SpaceOps-2025, paper 227}
}

@article{reiley1994,
  author  = {Reiley, D.~J. and Chipman, R.~A.},
  title   = {{Coating-induced wave-front aberrations: on-axis astigmatism and chromatic aberration in all-reflecting systems}},
  journal = {Applied Optics},
  volume  = {33},
  number  = {10},
  pages   = {2002--2012},
  year    = {1994},
  doi     = {10.1364/AO.33.002002}
}

@article{liang2001,
  author  = {Liang, C. and Descour, M.~R. and Sasian, J.~M. and Lerner, S.~A.},
  title   = {{Multilayer-coating-induced aberrations in extreme-ultraviolet lithography optics}},
  journal = {Applied Optics},
  volume  = {40},
  number  = {1},
  pages   = {129--135},
  year    = {2001},
  doi     = {10.1364/AO.40.000129}
}

@misc{wfc3ihb,
  author    = {Marinelli, M. and Dressel, L.},
  title     = {{Wide Field Camera 3 Instrument Handbook, Version 16.0}},
  publisher = {STScI},
  address   = {Baltimore},
  year      = {2024},
  note      = {\url{https://hst-docs.stsci.edu/wfc3ihb}}
}

@misc{tuttle2024,
  title={Ultraviolet Technology To Prepare For The Habitable Worlds Observatory}, 
  author={Sarah Tuttle and Mark Matsumura and David R. Ardila and Pin Chen and Michael Davis and Camden Ertley and Emily Farr and Brian Fleming and Kevin France and Cynthia Froning and Fabien Grisé and Erika Hamden and John Hennessy and Keri Hoadley and Stephan R. McCandliss and Drew M. Miles and Shouleh Nikzad and Manuel Quijada and Isu Ravi and Luis Rodriguez de Marcos and Paul Scowen and Oswald Siegmund and Carlos J. Vargas and Dmitry Vorobiev and Emily M. Witt},
  year={2024},
  eprint={2408.07242},
  archivePrefix={arXiv},
  primaryClass={astro-ph.IM},
  note={\url{https://arxiv.org/abs/2408.07242}} 
}

@ARTICLE{massey13,
       author = {{Massey}, Richard and {Hoekstra}, Henk and {Kitching}, Thomas and {Rhodes}, Jason and {Cropper}, Mark and {Amiaux}, J{\'e}r{\^o}me and {Harvey}, David and {Mellier}, Yannick and {Meneghetti}, Massimo and {Miller}, Lance and {Paulin-Henriksson}, St{\'e}phane and {Pires}, Sandrine and {Scaramella}, Roberto and {Schrabback}, Tim},
        title = "{Origins of weak lensing systematics, and requirements on future instrumentation (or knowledge of instrumentation)}",
      journal = {\mnras},
         year = 2013,
        month = feb,
       volume = {429},
       number = {1},
        pages = {661-678},
          doi = {10.1093/mnras/sts371},
archivePrefix = {arXiv},
       eprint = {1210.7690},
 primaryClass = {astro-ph.CO},
       adsurl = {https://ui.adsabs.harvard.edu/abs/2013MNRAS.429..661M}
}

@inproceedings{feinberg2024,
  author    = {Feinberg, L. and Ziemer, J. and Ansdell, M. and Crooke, J. and
               Dressing, C. and Mennesson, B. and O'Meara, J. and Pepper, J. and
               Roberge, A.},
  title     = {{The Habitable Worlds Observatory engineering view: status, plans, and opportunities}},
  booktitle = {Space Telescopes and Instrumentation 2024: Optical, Infrared, and Millimeter Wave},
  series    = {Proc. SPIE},
  volume    = {13092},
  pages     = {130921N},
  year      = {2024},
  doi       = {10.1117/12.3018328}
}

@article{feinberg2026,
  author  = {Feinberg, L.~D. and Sitarski, B.~N. and McElwain, M.~W. and others},
  title   = {{Habitable Worlds Observatory's concept and technology maturation: initial feasibility and trade space exploration}},
  journal = {ArXiv e-prints},
  year    = {2026},
  eprint  = {2601.11803},
  archivePrefix = {arXiv}
}

@article{liu2026,
  author  = {Liu, A. and Levine, M. and Noecker, C. and others},
  title   = {{Early architecture concepts for the Habitable Worlds Observatory: system design, modeling, and analysis}},
  journal = {Journal of Astronomical Telescopes, Instruments, and Systems},
  volume  = {12},
  pages   = {041017},
  year    = {2026},
  doi     = {10.1117/1.JATIS.12.4.041017},
  eprint  = {2602.11046},
  archivePrefix = {arXiv}
}

@article{akeson19,
  author  = {Akeson, R. and Armus, L. and Bachelet, E. and others},
  title   = {{The Wide Field Infrared Survey Telescope: 100 Hubbles for the 2020s}},
  journal = {ArXiv e-prints},
  year    = {2019},
  eprint  = {1902.05569},
  archivePrefix = {arXiv}
}

@ARTICLE{arenberg25,
       author = {{Arenberg}, Jonathan W.},
        title = "{Advancing the Concept Maturity Level of the Servicing Architecture for Habitable Worlds Observer}",
      journal = {arXiv e-prints},
         year = 2025,
        month = oct,
          eid = {arXiv:2510.03214},
        pages = {arXiv:2510.03214},
          doi = {10.48550/arXiv.2510.03214},
archivePrefix = {arXiv},
       eprint = {2510.03214},
 primaryClass = {astro-ph.IM},
       adsurl = {https://ui.adsabs.harvard.edu/abs/2025arXiv251003214A}
}

@inproceedings{skottfelt26,
author = {Jesper Skottfelt and others},
title = {{Study of a four-band far-UV to near-IR high-resolution imager for the Habitable Worlds Observatory}},
volume = {14145},
series = {},
booktitle = {Space Telescopes and Instrumentation 2026: Optical, Infrared, and Millimeter Wave},
editor = {Laura Coyle and Shuji Matsuura and Marshall Perrin},
organization = {International Society for Optics and Photonics},
publisher = {SPIE},
pages = {141451N},
year = {2026},
doi = {10.1117/12.3104851},
URL = {https://doi.org/10.1117/12.3104851}
}

@inproceedings{bolcar26,
author = {Matthew Bolcar and Feng Zhao and Paul Scowen and Lee Feinberg and Jeffrey Smith and Josh Abel},
title = {{The Habitable Worlds Observatory technology development plan}},
volume = {14145},
series = {},
booktitle = {Space Telescopes and Instrumentation 2026: Optical, Infrared, and Millimeter Wave},
editor = {Laura Coyle and Shuji Matsuura and Marshall Perrin},
organization = {International Society for Optics and Photonics},
publisher = {SPIE},
pages = {141451L},
year = {2026},
doi = {10.1117/12.3104860},
URL = {https://doi.org/10.1117/12.3104860}
}

@misc{rauer2024,
      title={The PLATO Mission}, 
      author={Heike Rauer and others},
      year={2024},
      eprint={2406.05447},
      archivePrefix={arXiv},
      primaryClass={astro-ph.IM},
      url={https://arxiv.org/abs/2406.05447}, 
}

\label{lastpage}
\renewcommand{\item}{\originalItem}
\end{document}